\documentclass[times, twocolumn]{aastex63}
\usepackage{CJK}
\usepackage{graphicx}
\usepackage{enumerate}
\usepackage{amssymb, amsmath}
\usepackage{natbib}
\usepackage{color}
\usepackage{ulem}
\usepackage{dblfnote}
\usepackage{appendix}
\usepackage{hyperref}
\usepackage[stable]{footmisc}
\usepackage{comment}
\usepackage{footnote}
\usepackage{lineno}
\usepackage{soul}

\usepackage[nodayofweek]{datetime}
\newdateformat{monthyearday}{%
 \THEYEAR\ \monthname[\THEMONTH] \twodigit{\THEDAY}}

\newcommand{\mj}{\ensuremath{\,M_{\rm J}}}

\newcommand{\alyr}{$\alpha$ Lyr}
\newcommand{\me}{M$_\oplus$}

\newcommand{\mum}{$\mu$m}

\begin{document}
\submitjournal{AAS Journals}
\shorttitle{JWST Vega}
\shortauthors{Beichman et al.}


\title{Searching for Planets Orbiting Vega with the James Webb Space Telescope}

\correspondingauthor{Charles Beichman,
chas\@ipac.caltech.edu}

\author[0000-0002-5627-5471]{Charles Beichman}
\affiliation{NASA Exoplanet Science Institute, IPAC, Pasadena, CA 91125}
\affiliation{Jet Propulsion Laboratory, California Institute of Technology, Pasadena, CA 91109}

\author[0000-0001-5966-837X]{Geoffrey Bryden}
\affiliation{Jet Propulsion Laboratory, California Institute of Technology, Pasadena, CA 91109}

\author[0000-0002-3414-784X]{Jorge Llop-Sayson}
\affiliation{Jet Propulsion Laboratory, California Institute of Technology, Pasadena, CA 91109}

\author[0000-0001-7591-2731]{Marie Ygouf}
\affiliation{Jet Propulsion Laboratory, California Institute of Technology, Pasadena, CA 91109}

\author[0000-0002-7162-8036]{Alexandra Greenbaum}
\affiliation{IPAC, California Institute of Technology, Pasadena, CA 91125}

\author[0000-0002-0834-6140]{Jarron Leisenring}
\affiliation{Steward Observatory, University of Arizona, Tucson, AZ, 85721}

\author[0000-0001-8612-3236]{Andras Gaspar}
\affiliation{Steward Observatory, University of Arizona, Tucson, AZ, 85721}

\author{John Krist}
\affiliation{Jet Propulsion Laboratory, California Institute of Technology, Pasadena, CA 91109}

\author[0000-0003-2303-6519]{George Rieke}
\affiliation{Steward Observatory, University of Arizona, Tucson, AZ, 85721}

\author[0000-0002-9977-8255]{Schuyler Wolff}
\affiliation{Steward Observatory, University of Arizona, Tucson, AZ, 85721}


\author[0000-0002-3532-5580]{Kate Su}
\affiliation{Space Science Institute, Boulder, CO 80301}

\author[0000-0003-0786-2140]{Klaus Hodapp}
\affiliation{University of Hawaii, Hilo, HI, 96720}


\author[0000-0003-1227-3084]{Michael Meyer}
\affiliation{University Of Michigan, Madison, MI}

\author{Doug Kelly}
\affiliation{Steward Observatory, University of Arizona, Tucson, AZ, 85721}


\author{Martha Boyer}
\affiliation{Space Telescope Science Institute, 3700 San Martin Drive, Baltimore, MD 21218, USA}

\author[0000-0002-6773-459X]{Doug Johnstone}
\affiliation{NRC Herzberg Astronomy and Astrophysics, 5071 West Saanich Rd, Victoria, BC, V9E 2E7, Canada}
\affiliation{Department of Physics and Astronomy, University of Victoria, Victoria, BC, V8P 5C2, Canada}

\author[0000-0001-9886-6934]{Scott Horner}
\affiliation{NASA Ames Research Center, Mountain View, CA, 94035}

\author[0000-0002-7893-6170]{Marcia Rieke}
\affiliation{Steward Observatory, University of Arizona, Tucson, AZ, 85721}

\begin{abstract}
The most prominent of the IRAS debris disk systems, $\alpha$ Lyrae (Vega), at a distance of 7.7 pc, has been observed by both the NIRCam and MIRI instruments on the James Webb Space Telescope (JWST). This paper describes NIRCam coronagraphic observations which have achieved F444W contrast levels of 3$\times10^{-7}$ at 1\arcsec\ (7.7 au), 1$\times10^{-7}$ at 2\arcsec\ (15 au) and few $\times 10^{-8}$ beyond 5\arcsec\ (38 au), corresponding to masses of $<$ 3, 2 and 0.5 \mj\ for a system age of 700 Myr. Two F444W objects are identified in the outer MIRI debris disk, around 48 au. One of these is detected by MIRI, appears to be extended and has a spectral energy distribution similar to those of distant extragalactic sources. The second one also appears extended in the NIRCam data suggestive of an extragalactic nature.The NIRCam limits within the inner disk (1\arcsec\ --10\arcsec) correspond to a model-dependent masses of 2$\sim$3 \mj. \citet{Su2024} argue that planets larger even 0.3\mj\ would disrupt the smooth disk structure seen at MIRI wavelengths. Eight additional objects are found within 60\arcsec\ of Vega, but none has astrometric properties or colors consistent with planet candidates. These observations reach a level consistent with expected Jeans Mass limits. Deeper observations achieving contrast levels $<10^{-8}$ outside of $\sim$4\arcsec\ and reaching masses below that of Saturn are possible, but may not reveal a large population of new objects.
\vspace{0.5 in}
\end{abstract}


\date{August 2023}

\section{Introduction\label{sec:intro}}

The star Vega, $\alpha$ Lyrae, is a bright (V=0.03 mag), nearby (7.68 pc), A0V star (Table~\ref{tab:star}). Once considered the touchstone for stellar calibration, Vega gave its name to a new phenomenon when initial calibration observations by the Infrared Astronomical Satellite (IRAS) revealed a strong excess at long wavelengths attributable to a ring of cold dust emitting at 25-100 \mum\ \citep{Aumann1984}. Subsequently, the IRAS, Spitzer, WISE and Herschel telescopes identified hundreds of stars showing infrared emission from analogs of our own Kuiper Belt (cold dust at 10s of au, warm dust - asteroid belt analogs - at a few au, or both \citep{Wyatt2008,Su2013}. Approximately 20\% of mature main sequence stars show some form of far-infrared excess at currently detectable levels of a few tens of times the brightness of our own Kuiper Belt \citep{Wyatt2018}. 

The most notable nearby examples of the ``Vega Phenomenon" from IRAS include Vega, Fomalhaut, $\beta$ Pictoris, and $\epsilon$ Eri \citep{Gillett1986}. It was soon realized that the circumstellar material consists of micron and sub-micron sized dust particles resulting from collisions of larger bodies, or planetesimals. This ``debris disk" material is heated by the central star to emit in the infrared. In a few cases, notably that of $\beta$ Pictoris \citep{Smith1984} and Fomalhaut \citep{Kalas2005}, coronagraphic images in the visible revealed a corresponding disk or ring of material in scattered light.

\begin{deluxetable*}{llll}[t!]
\tabletypesize{\scriptsize}
\tablewidth{0pt}
\tablecaption{Properties of the Host Star Vega$^*$\label{tab:star}
}
\tablehead{
\colhead{Property} & \colhead{Value}& \colhead{Units}& \colhead{Comments} }
\startdata
Spectral Type & A0V & &\\
T$_{\rm eff}$ &9500-10059 & K &\citet{Yoon2010, Su2013}\\
Mass &2.15$^{+0.10}_{-0.15}$ & M$_\odot$&\citet{Monnier2012}\\
Luminosity & 56.0& L$_\odot$&\citet{Yoon2010, Su2013}\\
Age$^a$ &700$^{+150}_{-75}$ &Myr & \citet{Monnier2012}\\
$[$Fe/H$]$ &$-$0.56& dex &\citet{Baines2018}\\
log(g)&4.02$\pm$0.014&cgs &\citet{Yoon2010}\\	 
inclination ($i$)&5$\pm$2$^o$&&\citet{Monnier2012}\\	
R.A.\ (Eq 2000; Ep 2000)&18$^h$36$^m$56.34$^s$ & & \citet{Hipparcos2007} \\
Dec.\ (Eq 2000; Ep 2000)&$+$38$^o$47$^\prime$01.28\arcsec\ & & \citet{Hipparcos2007} \\
R.A.$^*$\ (Eq 2000; Ep 2023.633)&18$^h$36$^m$56.734$^s$ & & \\
Dec.$^*$\ (Eq 2000; Ep 2023.633)&$+$38$^o$47$^\prime$8.121\arcsec\ & & \\
Distance & 7.68$\pm$0.02 & pc & \citet{Hipparcos2007}\\
Proper Motion ($\mu_\alpha,\mu_\delta$)&(200.94, 286.23) &mas/yr &\citet{Hipparcos2007}\\
F210M & 0 mag (688 Jy) & &See text\\
F444W & 0 mag (184 Jy) & &See text\\
\enddata
\tablecomments{The physical properties of Vega are hard to summarize simply due to the star's rapid rotation and oblateness as seen from almost pole-on \citep{Yoon2008,Monnier2012}. This leads to a range of effective temperatures and other parameters depending on orientation and rotation. Representative values are given. The age of 700 Myr is older than previous models due to these effects as discussed in the text. $^*$As observed from vantage point of JWST L2 orbit at the epoch of observation, 2023-08-19.}
\end{deluxetable*}

\begin{deluxetable*}{llllllllll}
\tablewidth{0pt}
\tablecaption{NIRCam Imaging Observing Parameters (PID:\#1193)\label{tab:exposures}
}
\tablehead{\colhead{Visit}&
\colhead{Target} & \colhead{Mask}&\colhead{Filter}&
\colhead{Subarray}
&\colhead{Readout} & \colhead{Groups/Int} & \colhead{Ints/Exp}& \colhead{Dithers} & \colhead{Total Time (sec)}}
\startdata
34&$\alpha$ Lyr (Roll \#1) &M430R&F210M \& F444W&FULL &RAPID &3&25&1 &1063 \\
35&$\alpha$ Lyr (Roll \#1) &M430R&F210M \& F444W&SUB320 &RAPID &8&150&1 &1446 \\
36&$\alpha$ Lyr (Roll \#2) &M430R&F210M \& F444W&SUB320 &RAPID &8&150&1 &1446 \\
37&$\alpha$ Lyr (Roll \#2) &M430R&F210M \& F444W&FULL &RAPID &3&25&1 &1063 \\\hline
38&$\alpha$ Cyg (5-pt Dither) &M430R&F210M \& F444W&SUB320 &RAPID &8&100&5 &4820 \\
39&$\alpha$ Cyg (5-pt Dither) &M430R&F210M \& F444W&FULL &RAPID &3&8&5 &1664 \\
\enddata
\tablecomments{Observations were obtained on 2023-Aug-19 at 05:35-11:50 UTC}
\end{deluxetable*}

\begin{figure*}[t!]
\centering
\includegraphics[width=0.55\textwidth]{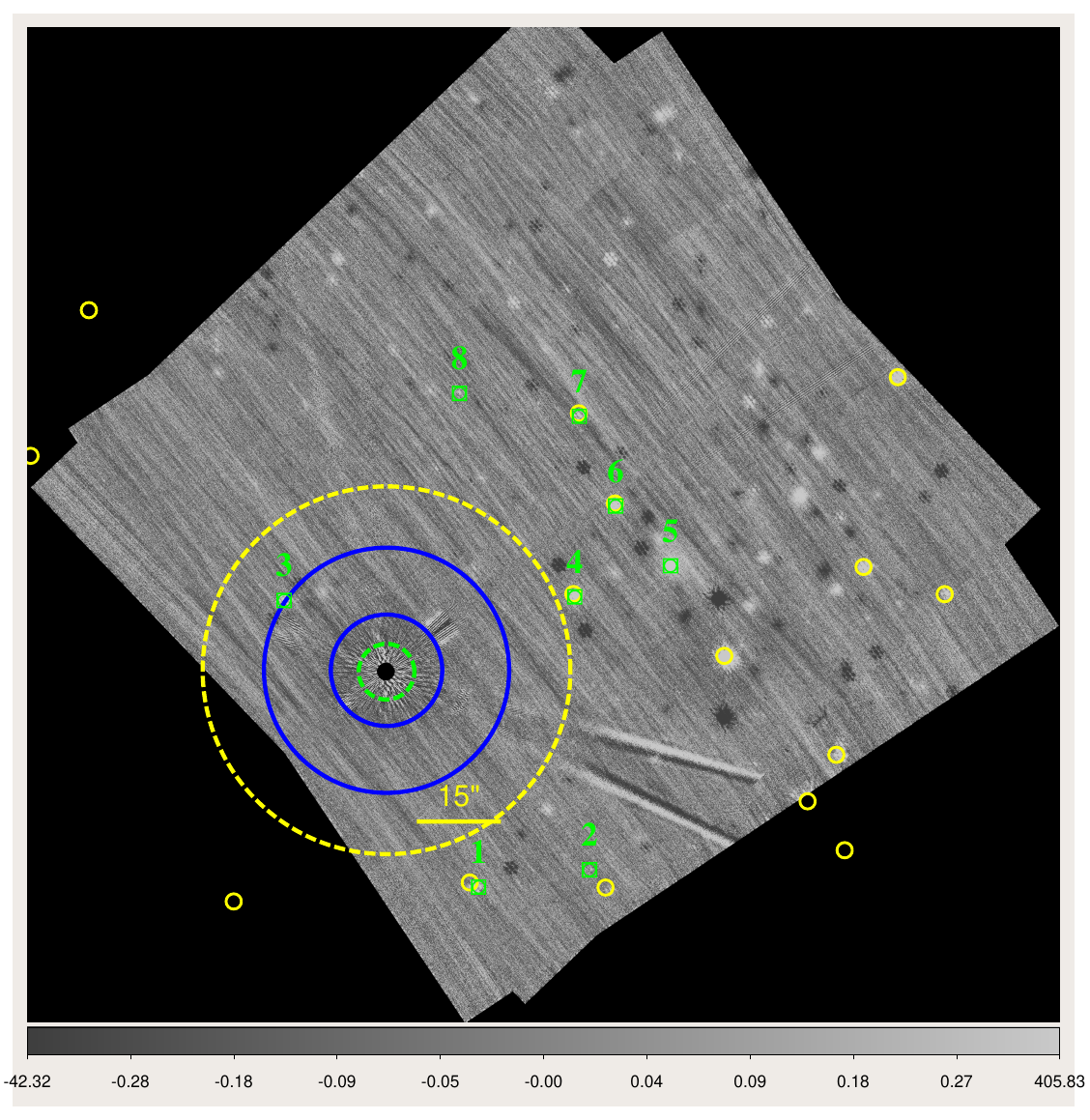}\\
\includegraphics[width=0.55\textwidth]{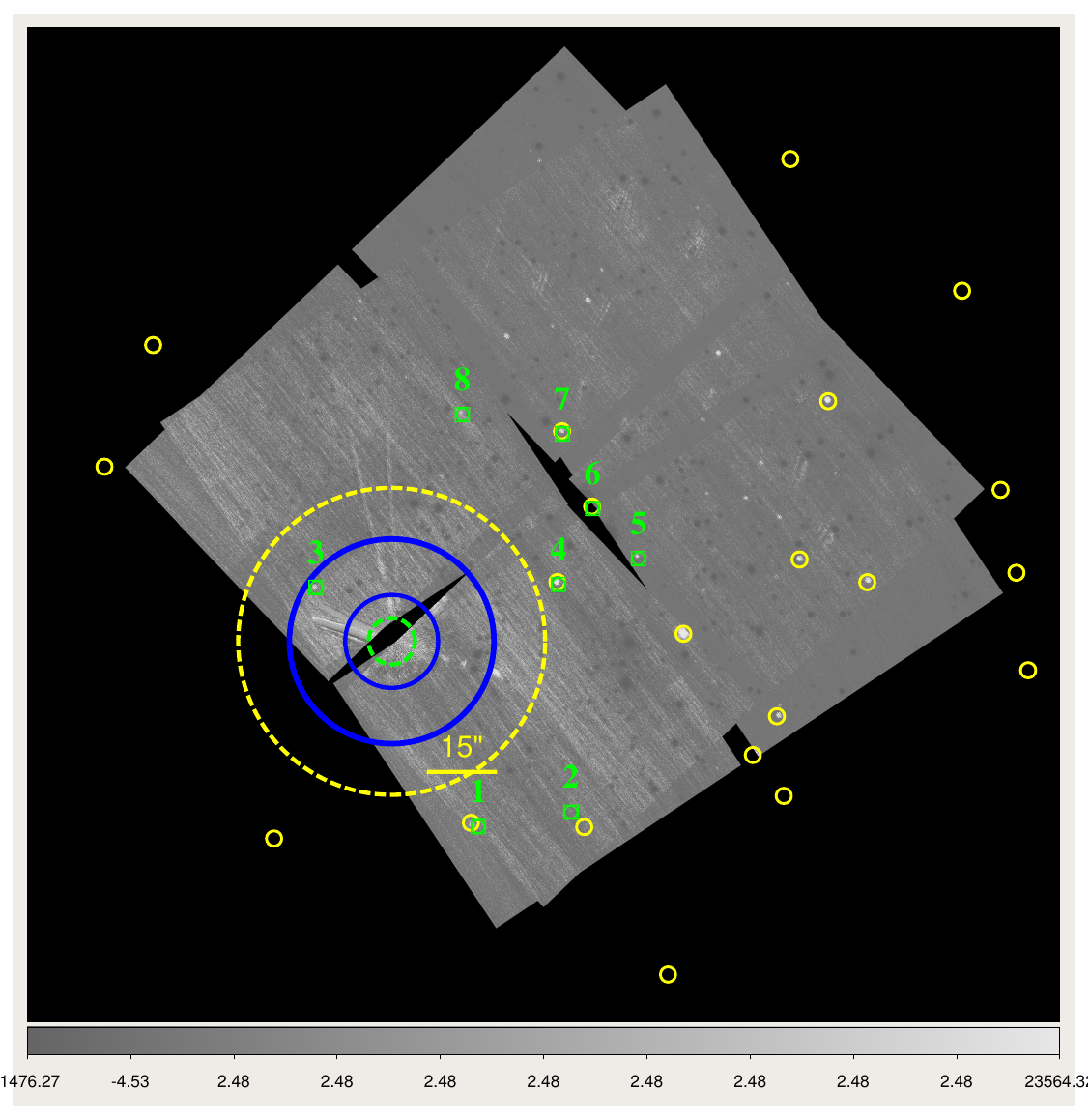}
\caption{top) The full-frame F444W image generated using ADI imaging. The dark holes bracketing each bright source are an expected artifact of the roll subtraction and serve as a check for the reality of each object. The brightest 8 F444W sources within 60\arcsec\ of Vega and not obviously extended are indicated (Table~\ref{tab:sources2}). In the figure North is up and East to the left. The large circles delineate features seen in the MIRI data \citep{Su2024}: the dotted yellow circle denotes the outer extent of the halo seen in the F2550W image (33\arcsec); the blue circles denote the extent of the Kuiper Belt Analog seen in both MIRI and ALMA (10\arcsec-22\arcsec); an inner disk extends inward to the star while the dotted green ring marks the location of a dip in the dust distribution seen around 5\arcsec-10\arcsec. Small yellow circles denote Gaia sources brighter than Gmag$<$18 mag. bottom) Full-frame image for F210M, with PSF subtraction performed with ADI+RDI. The many dark spots represent sources in the field of the reference image. Intensity units are in MJy sr$^{-1}$. \label{fig:NIRCamSources}}
\end{figure*}

The evidence for solid material orbiting main sequence stars immediately led to the supposition that this material represented the remnants of the planet formation process. There is growing evidence for a strong correlation between the presence of planets and the presence of a debris disk \citep{Meshkat2017}. Indeed, direct imaging has revealed Jovian mass planets among a number of the most prominent debris disk systems: $\beta$ Pic \citep{Lagrange2010}, $\epsilon$ Eri \citep{Mawet2019} and HR 8799 \citep{Marois2008}. However, the object claimed to be a planet orbiting Fomalhaut \citep{Kalas2008} is likely due to a collision between two comets \citep{Lawler2015,Gaspar2020,Gaspar2023,Ygouf2023}.

In the case of Vega, an analysis of data from Spitzer and Herschel suggest the existence of warm $\sim$170 K dust (an asteroid belt analog) at a separation of 10-14 au (1.3\arcsec-2\arcsec) from the star and cold $\sim$50 K dust (a Kuiper Belt analog) at $\sim$110 au (beyond $\sim$10\arcsec) \citep{Su2013}. ALMA imaging at 1.34 mm shows a bright face-on belt starting sharply at 60-80 au ($\sim$10\arcsec) and extending outward to 150-200 au (20-26\arcsec) \citep{Matra2020}. These authors suggest that the sharp inner edge of the dust belt seen by ALMA argues in favor of a shepherding planet with mass $>$6\me\ around 70 au. There is also a poorly understood population of extremely hot grains at the dust sublimation temperature of 1500-2000 K, seen in ground based interferometry \citep{Ciardi2001,Absil2006}.

 Most recently, \citet{Su2024} reported MIRI observations of Vega's disk which provide much greater detail than earlier observations. Briefly summarized, the MIRI images reveal a smooth halo extending out to 33\arcsec\ (250 au); a broad Kuiper-Belt-analog ring extending from $\sim$10\arcsec\ to 22\arcsec\ ($\sim$78 to 70 au) that coincides with the parent body system detected with ALMA at 1.3 mm; a gap in the in the disk around 60 au; an inner disk extending to within very close the star $<$0.6\arcsec\ (3-5 au) of the star; and a dip in surface brightness of the inner disk from $\sim$5\arcsec\ to 10\arcsec\ ($\sim$40–78 au). Overall, the inferred dust distribution is very smooth and argues against the presence of planets more massive than $> 0.3$ \mj\ orbiting the star outside of about 10 au.

To date, searches for a planet orbiting Vega have not borne fruit with the following mass limits assuming ages of 400-800 Myr: 4 - 35\mj\ at orbital radii of 170–260 au from Palomar \citep{Metchev2003}; 5-10 \mj\ at 10-70 au set with the MMT \citep{Heinze2008}; 2-4 \mj\ at separations of 100-200 au set with Spitzer \citep{Janson2015}; 30-40 \mj\ within 15 au from Palomar \citep{Meshkat2018}; and 3 \mj\ between 1-12 au \citep{Ren2023} based on Vector Vortex coronagraphy from Keck. Most recently, \citet{Hurt2021} have used a decade of Precision Radial Velocity (PRV) data to identify the possible presence of a 20 \me/$sin(i)$ planet in a 2.43 day orbit. The pole-on orientation of \alyr\ means that the planet could have a much higher mass. Their mass limits also rule out planets with 1 \mj/$sin(i)$ at 1 au, 5 \mj/$sin(i)$ at 10 au and 13 \mj/$sin(i)$ at 15 au.

 Outside of 1\arcsec\ ($\sim$ 7.7 AU at Vega) JWST offers major gains compared with ground-based telescopes, reaching mass levels as low as a Saturn mass \citep{Carter2023,Ygouf2023}. A primary goal of the GTO program \#1193 is a search for planets orbiting three of the most prominent debris systems mentioned above: Vega, Fomalhaut and $\epsilon$ Eri. This paper reports on the search for planets orbiting Vega that achieves a sensitivity of $<$1 \mj\ at separations outside 4\arcsec\ (30 au).
 
Finally, we note that Vega's role as a touchstone system was further revised when it was realized that it was an oblate, rapidly rotating star seen almost pole-on \citep{Peterson2006}. Updated models have led to a revision of the star's age from an initial value of 450$\pm$50 Myr \citep{Yoon2010} to an older age of 700$^{+75}_{-150}$ Myr based on interferometric measurements \citep{Monnier2012}. The age increase means that the conversion of planet brightness (or upper limit) will result in somewhat higher masses than for the younger age. We adopt the 700 Myr age but note in some places the effect of assuming a younger age.

 {\color{black} Section $\S$\ref{sec:observations} describes the observations. $\S$\ref{sec:reduce} describes the data reduction steps, presents images close to the star ($<$ 10\arcsec) and over a much wider field, describes the limiting contrast level ($\S$\ref{sec:contrast}), lists a number of sources detected in the images, and details a search for scattered disk emission ($\S$\ref{sec:disk}). $\S$\ref{sec:results} assesses the potential of the point sources as exoplanet candidates and discusses the surface brightness limits in terms of disk properties. } $\S$\ref{sec:discuss} discusses both the sub-Jovian mass limits achieved here in the context of planet formation mechanisms and the limits on scattered light. $\S$\ref{sec:conclude} gives our conclusions.

\section{Observations}\label{sec:observations}

JWST observed Vega through the round M430R coronagraphic mask simultaneously at two wavelengths (F210M and F444W), one each in the Long and Short wavelengths arms of NIRCam {\color{black}(\citealt{Rieke2023,Girard2022}; }Table~\ref{tab:exposures}). The observations were obtained at two roll angles separated by 7\arcdeg. Because of the brightness of Vega, we split the observations into: 1) a set of subarray (SUB320) exposures optimized to avoid saturation close to the star but with a limited field of view extending only $\pm$10\arcsec; and 2) a set of observations in full array mode providing {\color{black} coverage of} the full extent of the debris disk out beyond $\pm$30\arcsec\ \citep{Su2013,Matra2020}. 

The exposure time at F444W was chosen to search for planets down to $<$ 1 \mj\ masses at 4\arcsec\ assuming a 5 nm wavefront drift and using representative models for giant planets, e.g. \citet{Spiegel2012}. At this separation, we should be able to detect a 1 \mj\ planet with a SNR of about 5. The simultaneous F210M observations are used to identify and reject (in a preliminary manner) background stars or extragalactic objects based on their [F210M]-[F444W] color. Final confirmation of the association of a new source with Vega would require a second astrometric observation at a later epoch.

We adopted the A2Ia star Deneb ($\alpha$ Cyg, HD 197345; K$_s$=0.88 mag) as a Point Spread Function (PSF) reference. The star is 24\arcdeg\ deg away from Vega, but for the observing date in question 2023-Aug-19, the change in solar offset angle between the two stars is only 9.3\arcdeg\ which reduced the effects of changing thermal environment on the JWST telescope \citep{Perrin2018}. Vega and Deneb have similar but not identical spectral types (A0V and A2I, respectively which leads to small differences in near-IR colors, K-L=-0.03 mag and 0.11 mag, respectively \citep{Johnson1966}. According to JWST documentation this color difference will have little or no effect on the achievable F444W contrast within $<$2\arcsec\ \citep{JDoxColor} and no effect at larger separations. 

Table~\ref{tab:exposures} describes the NIRCam observing parameters. We used the 5-POINT dither pattern for the Deneb observations to increase the diversity in the PSF for post-processing and thus to increase the contrast gain at close separations. For the SUB320 observations where PSF subtraction is critical, we maintained a similar SNR per frame for both targets. For the FULL frame images we relied primarily on Angular Differential Imaging (ADI) so the PSF observations per dither position were shorter than for Vega.

\section{Data Reduction and Post Processing\label{sec:reduce}}

The pipeline processing and and post-processing steps closely follow the procedures described in \citet{Ygouf2023} and are summarized below.

\subsection{Pipeline Processing} 

The full set of images (summarized in Table~\ref{tab:exposures})
was processed using the {\tt JWST} pipeline version \rm{2023\_3b}, calibration version
\rm{1.9.6}, CRDS context for reference files \rm{jwst\_1202.pmap}, 
photometry reference file \rm{jwst\_nircam\_photom\_0157.fits}, and
distortion reference file \rm{jwst\_nircam\_distortion\_0173.asdf}. 
The dataset can be obtained at: \dataset[https://doi.org/10.17909/76j1-4g22]{https://doi.org/10.17909/76j1-4g22}.

Standard pipeline processing \citep{jwst2022} was used with some modifications: 1) dark current corrections are not well characterized for subarray observations and were not used; 2) the {\tt SpaceKLIP} package \citep{Kammerer2022} was used for ramp fitting which significantly improved the noise floor in the subarray images and reduced 1/f noise; and 3) measurements with only a single group before saturation were accepted to reduce saturation effects in the full array images.

\subsection{Bad Pixel Rejection}

As described in \citet{Ygouf2023}, we utilized flagging less conservative than the default, i.e., ({\tt n\_pix\_grow\_sat} set to 0, rather than 1). For identification of truly bad pixels, we used the pipeline {\tt DQ} flags: any pixels flagged as {\tt DO\_NOT\_USE}, e.g.\ dead pixels, those without a linearity correction, etc., were set to {\tt NaN}. 5-$\sigma$ outliers -- temporally within sub-exposures or spatially within a 5x5 box -- were also rejected. Additional bad pixels which became apparent following PSF subtraction (\S\ref{PSFsubtraction}) were similarly rejected.

\subsection{Point Spread Function (PSF) Subtraction}\label{PSFsubtraction}

Following the basic data reduction for individual images, we used two approaches to post processing: a combination of Angular Difference Imaging (ADI) and Reference star Differential Imaging (RDI) or ADI alone. 

For the classical RDI+ADI, we first created a reference PSF from the nearby star Deneb (HD 197345), shifting and coadding its five dithered observations together to maximize SNR. We scaled and shifted the reference star PSF to align with the target at Roll~1 and Roll~2 independently before performing the PSF subtraction. 
For the classical ADI, we subtracted the two rolls from one another after applying the corresponding shift and data centering. In both RDI+ADI and ADI approaches, the last step after PSF subtraction was to orient both subtracted rolls to the North before coadding them resulting in a negative-positive-negative pattern for sources that are present in both telescope angles. 

While the classical PSF subtraction performs well at larger distances from the target star where the noise is limited by the instrument sensitivity, at close separations residual starlight speckles are the dominant limitation to the detection of point sources. Principal Component Analysis (PCA) analysis is preferred for cleaning the inner speckle field within $\sim$1.5\arcsec. In this speckle dominated region we applied ADI combined with RDI, using a PCA-based algorithm \citep{Amara2012} via Karhunen Lo\'eve Image Projection \citep[KLIP;][]{Soummer2012}. We used the open source Python package \texttt{pyKLIP} \citep{Wang2015}, which provides routines for cleaning the images, calculating detection limits, and quantifying the uncertainty in the flux of any detected sources.

\begin{figure*}[t!]
\centering
\includegraphics[width=0.46\textwidth]{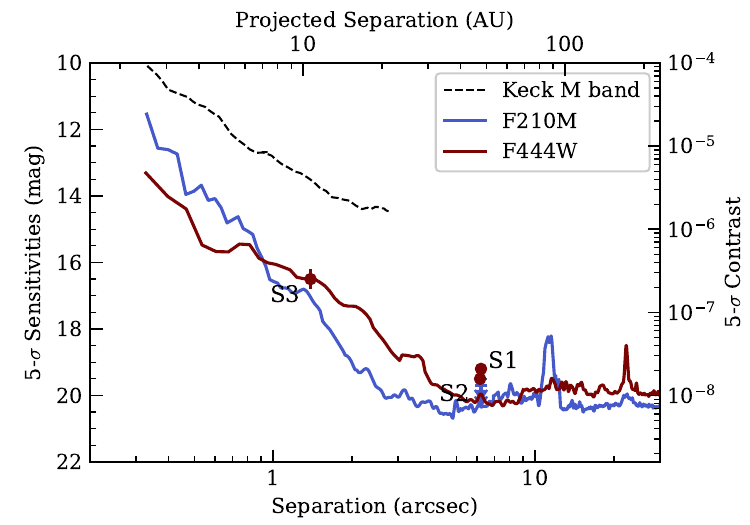} 
\hspace{0.2in}
\includegraphics[width=0.46\textwidth]{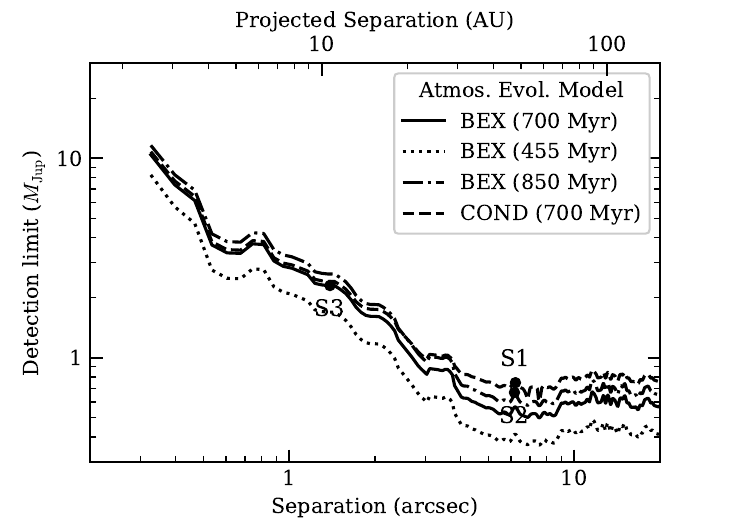} 
\caption{
Left:
Our detection limits are shown for each filter (F210M and F444W) as a function of separation from the parent star, in terms of contrast ratio (right-side axis) and apparent magnitude (left-side axis).
The drop in detectability at $\sim$9\arcsec corresponds to the boundary between subarray and full-frame imaging.
The limits are given as 5-$\sigma$.
The pre-JWST contrast limit at M band from Keck \citep{Ren2023} is shown as a dashed line for comparison; the JWST detections limits are a factor of $\sim$10 better within $\sim$2\arcsec, and improve even further at larger separations.
Right: the flux sensitivities on the left are translated to 5-$\sigma$ detection limits in terms of planet mass. 
For a nominal system age of 700 Myr, two models of atmosphere evolution are considered -- BEX-HELIOS \citep[solid line;][]{Linder2019} and Ames-COND \citep[dashed line;][]{baraffe03} and -- with similar results from both. A younger age (e.g.\ 450 Myr from \citet{Yoon2010}) would correspond to planet masses lower by $\sim$30\% (dotted line), while the upper limit from \citet{Monnier2012} (850 Myr) would result in slightly higher masses (dot-dash).
\label{fig:detectionLimits}}
\end{figure*}

The KLIP reduction was done using all available images (i.e. with 6 KL-modes), with the full array mode data cropped and centered to match the subarray mode data. Only the full array dataset was used to produce the PCA full-frame reductions. The reference star data was used for PSF subtraction. Use of the 5-POINT-SMALL-GRID dither pattern mitigated any misalignment between the star and coronagraph focal plane mask. The KLIP reduction in the inner region, in which the noise is dominated by the residual speckles from Vega, was done using a set of annuli with a width of 6 pixels, and each annulus was divided in 4 subsections. 

Figures~\ref{fig:NIRCamSources} show the results of the classical PSF and PCA subtractions. The presence of the expected negative-positive-negative image pattern is a good indication that a candidate object is real. Point source extraction is discussed below.

\begin{figure*}[t!]
\centering
 \includegraphics[width=1\textwidth]{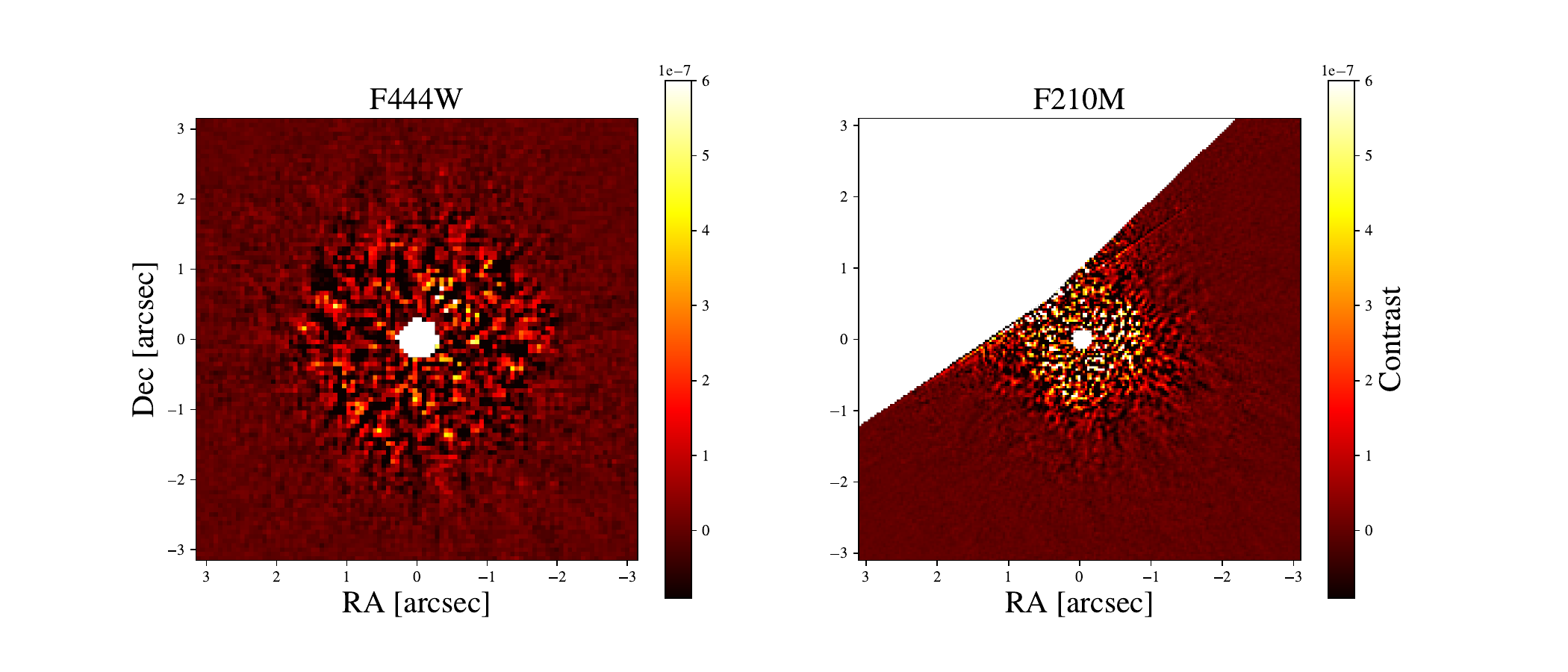}
\caption{PSF-subtracted images of the subarray data. The bright residuals are residuals from the imperfectly subtracted speckles. The North-East part of the F210M data (\textit{right}) is missing due to the PSF falling close to the SW detector gap, i.e. the edge of NIRCam detector A2.
\label{fig:SUBimages}}
\end{figure*}

\begin{deluxetable*}{llclllll}[t!]
\tablewidth{0pt}
\tablecaption{NIRCam Sources Found Close to Vega\label{tab:sources}}
\tablehead{&\colhead{Offset}&
\colhead{Separation} & \colhead{F444W contrast}&\colhead{F$_\nu$(F210M)}&
\colhead{F$\nu$(F444W)}&\colhead{F$\nu$(F1550W)} & \colhead{F$\nu$(F2550W)}\\
\colhead{ID}&\colhead{$\Delta \alpha,\Delta\delta$\, (\arcsec)} & \colhead{(AU)}&\colhead{($\times10^{-8})$}&
\colhead{($\mu$Jy)/(Vega mag)} &
\colhead{($\mu$Jy)/(Vega mag)} &
\colhead{($\mu$Jy)/(Vega mag)} &
\colhead{($\mu$Jy)/(Vega mag)}}
\startdata
S1$^*$&($-$5.52,$+$2.86)&48&$2.2\pm0.35$&$<$18.5 (5$\sigma$)&3.7$\pm$0.6&23$\pm$5&20$\pm$5\\
&$\pm$0.013\arcsec&&&$>$19.&19.2$\pm$0.15&14.6$\pm$0.2&13.7$\pm$0.4\\
S2$^*$&($+$1.49,$+$5.99)&47&$1.7\pm0.3$&$<$10&2.9$\pm$0.5& $<$35 ($3\sigma$)& $<$30 ($3\sigma$)\\
&$\pm$0.013\arcsec&&&$>$19.7&19.5$\pm$0.15& &\\
\hline
\enddata
\tablecomments{$^*$Appears to be extended at F444W. Positions with respect to ($\alpha\, , \delta$; Eq 2000; Ep 2023.633)= 18$^h$36$^m$56.734$^s$ $+$38$^o$47$^\prime$8.121\arcsec.
}
\end{deluxetable*}


\subsection{Contrast Calibration\label{sec:contrast}}

The contrast limits reported in this work are obtained by normalizing the flux to a synthetic peak flux. We estimated Vega's flux density in the NIRCam bands by convolving a Kurucz model of 9500 K and \textit{log g}=4.0 with the JWST bandpasses to obtain 688 Jy in the F210M filter and 184 Jy at F444W assuming V=0.03 mag \citep{Johnson1966}. To estimate the peak flux of the instrument's off-axis coronagraphic PSF we simulated this PSF using \texttt{WebbPSF} \citep{Perrin2014}. Measured fluxes in the NIRCam images are divided by these stellar fluxes to obtain contrast ratios.

The 5-$\sigma$ contrast curves (Figure~\ref{fig:detectionLimits}) are obtained using \texttt{pyKLIP}. The noise is computed in an azimuthal annulus at each separation, and we use a Gaussian cross correlation to remove high frequency noise. The contrast is calculated using the normalization peak value for the target star. We corrected for algorithmic throughput losses by injecting and retrieving fake sources at different separations. The contrast is also corrected for small sample statistics \citep{Mawet2014} at the closest angular separations. 
At separations closer than 2\arcsec\ the contrast is limited 
by the residuals from the PSF subtraction methods ($\sim3\times 10^{-7}$ at 1\arcsec, $\sim 1 \times 10^{-7}$ at 2\arcsec), and further than 2\arcsec\ the performance is limited by the background level ($\sim 19$\,mag in F444W), consistent with the expectations of the instrument given the exposure time (Figure~\ref{fig:SUBimages}). 
Fig~\ref{fig:detectionLimits}b converts the sensitivity curves into detection limits in Jupiter masses appropriate to Vega's age and distance ($< 1$ \mj\ beyond 2\arcsec; \citet{Linder2019,Baraffe2003}).

Losses due to the coronagraphic mask are taken into account in both the contrast curves and in the reported fluxes for any detected sources (the following section).

\begin{figure*}[t!]
\centering
\includegraphics[width=0.6\textwidth]{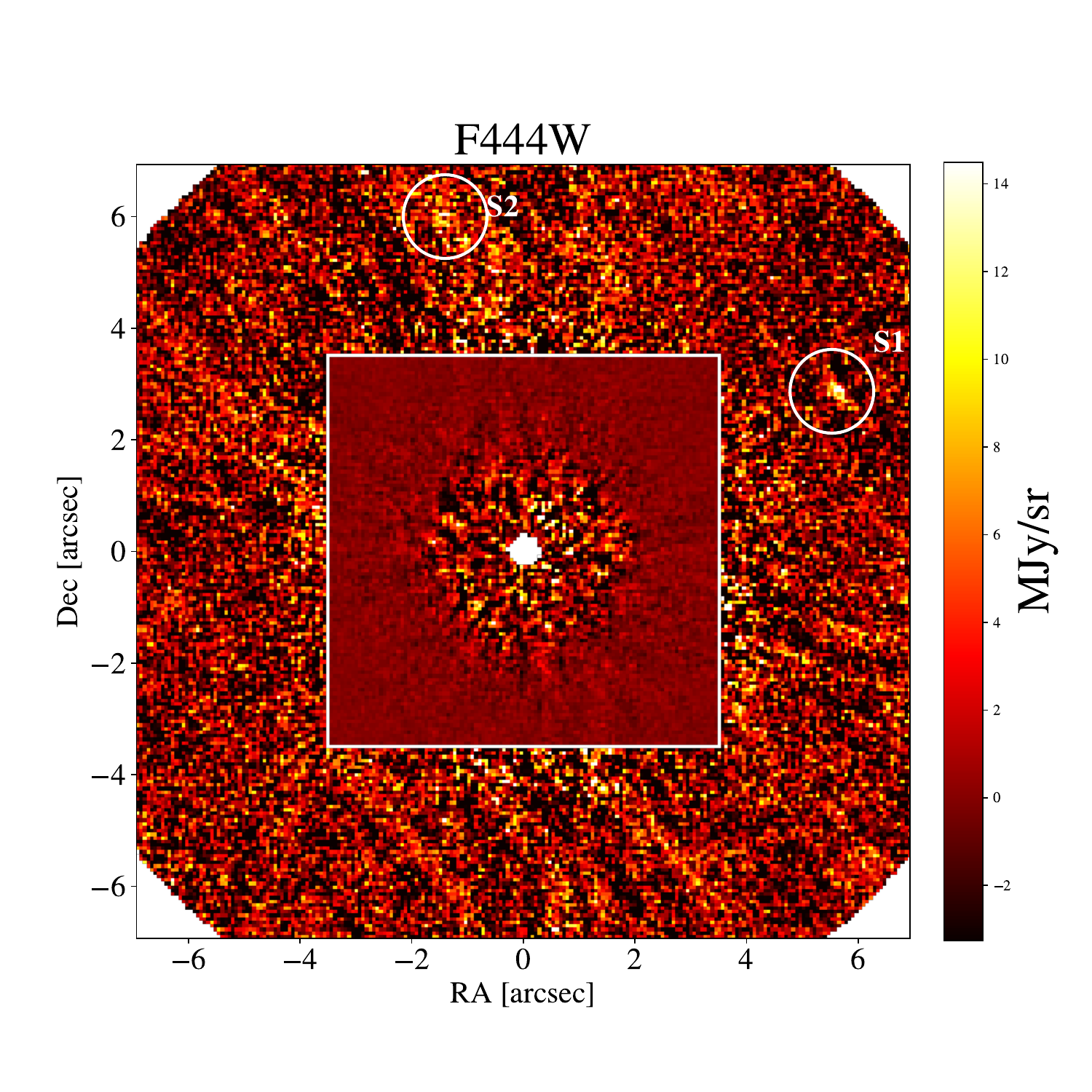} 
\caption{ NIRCam sub-array PCA reduction at F444W shows 2 sources: S1 and S2 are both extended. Two different stretches of the reduced image are superimposed to emphasize the sources; the colorbar is applicable to the inner region only. Details of the analysis are presented in the Appendix. \label{fig:schematic}}
\end{figure*}

\begin{figure*}[t!]
\centering
\includegraphics[width=0.7\textwidth]{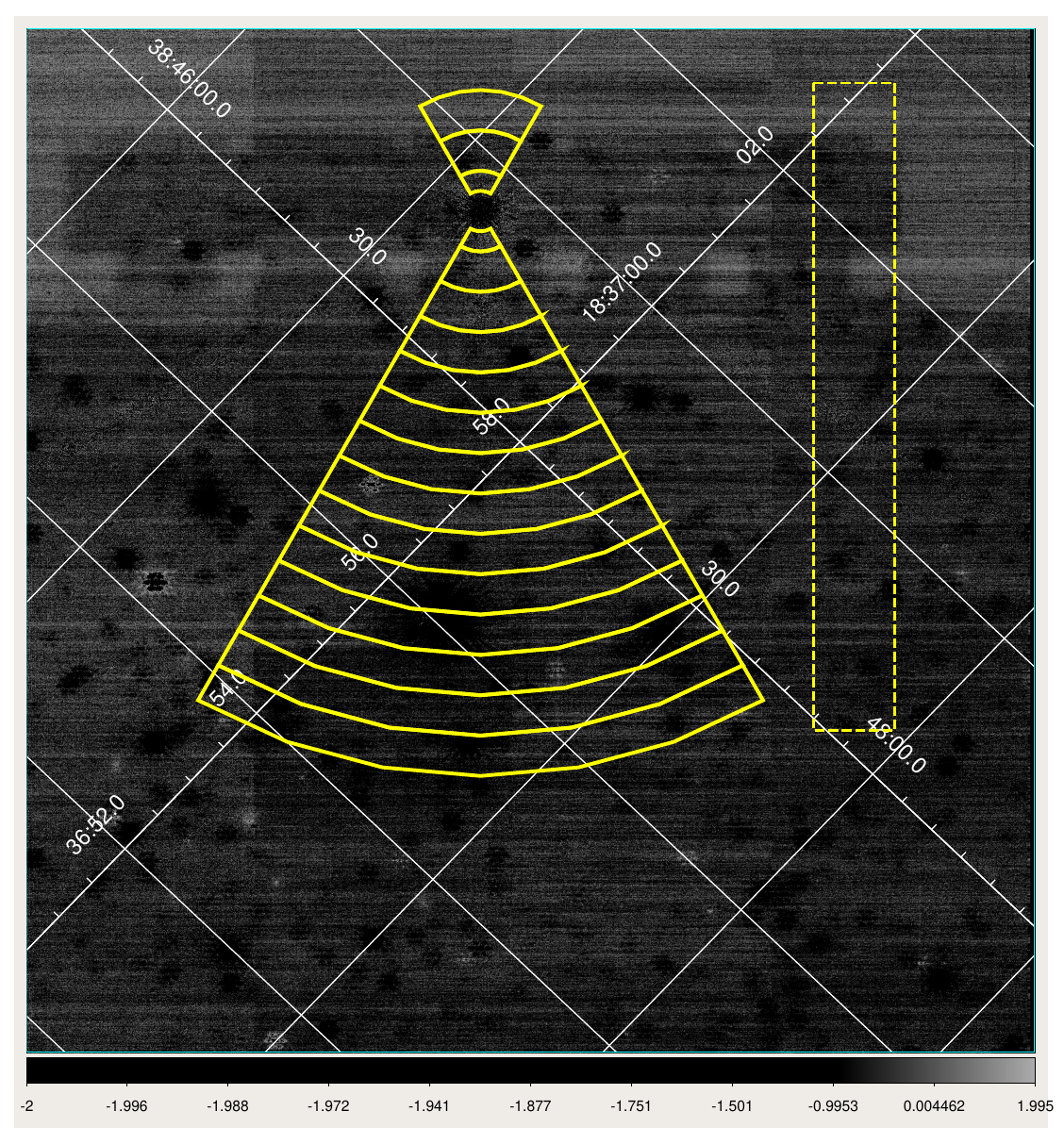}
\includegraphics[width=0.7\textwidth]{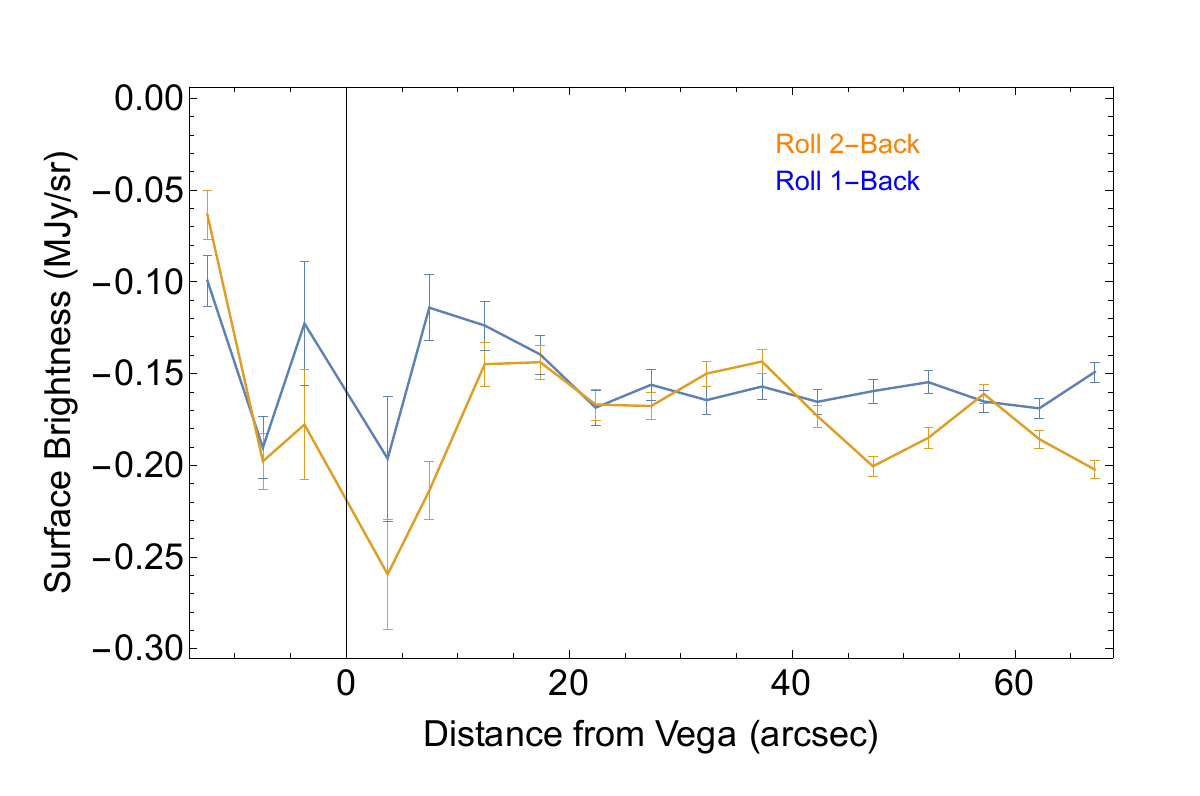}
\caption{top) Simple RDI subtraction of the F444W images of Vega and Deneb to search for disk emission. Arc-shaped annuli laying between the major Vega diffraction spikes were examined for disk emission as a function of distance from Vega. The dotted box denotes a region used as a "background" region to measure emission away from the annuli. bottom) The blue and orange lines denote emission within the partial annuli as a function of separation from Vega minus the "background" level in a vertical band at the right side of the image. The feature at $\sim$10\arcsec\ is an artifact visible in the image and in the background trace and is due to the edge of the coronagraphic optical mount. \label{fig:VegaDisk}}
\end{figure*}

The performance of filter F210M within 1\arcsec~is significantly worse with respect to F444W (see Fig~\ref{fig:detectionLimits}a). We find that the variability of the speckle pattern close to the center of the star is visibly worse for the F210M dataset, in other words, the speckles change more frame-to-frame between science images and reference images. Changes in the speckle pattern between the science target and the reference frames hinder the PSF subtraction and could explain the degraded performance of F210M close-in. 

\begin{figure*}[t!]
\centering
\includegraphics[width=0.8\textwidth]{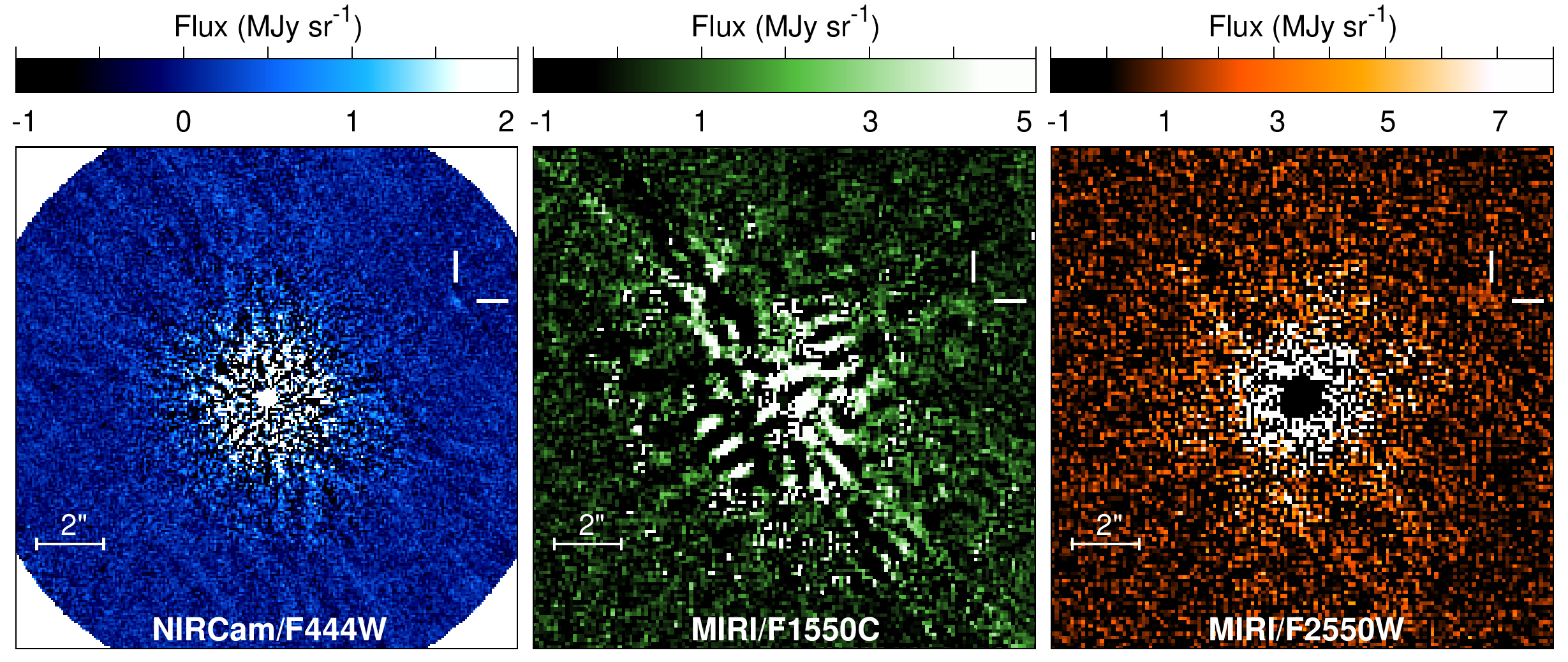} 
\caption{left) NIRCam sub-array PCA reduction at F444W showing the source $S1$ at $\sim$2 o'clock. center and right) MIRI observations optimized for point source detection identifies the same source at F1550 and F2550W, strongly suggesting this object is extragalactic in nature \citep{Su2024}. In all three images intensity units are MJy sr$^{-1}$. \label{fig:NIRCamMIRI}}
\end{figure*}

\begin{deluxetable*}{lllllc}[t!]
\tablewidth{0pt}
\tablecaption{Full Frame Sources Within 60\arcsec\ of Vega\label{tab:sources2}}
\tablehead{\colhead{ID}
&\colhead{RA$^1$}&
\colhead{Dec$^1$} & \colhead{F210M}&\colhead{F444W}&\colhead{[F210M]-[F444W]$^2$}\\
&\colhead{(Deg., Epoch=2023.63)} & \colhead{(Deg., Epoch=2023.63)} &\colhead{($\mu$Jy)/(Vega mag)}&\colhead{($\mu$Jy)/(Vega mag)}&\colhead{(mag)}}-
\startdata
1$^{3,4}$&279.230674 (55.36$^s$) & 38.774867 (46$^\prime$29.52\arcsec)&403.0$\pm $0.40 (14.15)&102.0$\pm$1.0 (15.64)&-0.06\\
2&279.223232 (53.58$^s$) & 38.775728 (46$^\prime$32.62\arcsec)
&19.5$\pm $0.12 (17.44)&5.0$\pm$0.8 (18.91)&-0.05\\
3$^3$&279.242866 (58.29$^s$) & 38.789151 (47$^\prime$20.94\arcsec)&309.0$\pm $0.59 (14.44)&67.0$\pm$0.8 (16.10)&-0.23\\
4$^3$&279.224276 (53.83$^s$) & 38.789329 (47$^\prime$21.58\arcsec)&1,378.6$\pm $0.97 (12.81)&419.4$\pm$0.9 (14.11)&0.14\\
5&279.218097 (52.34$^s$) & 38.790840 (47$^\prime$27.03\arcsec)
&76.6$\pm $0.23 (15.95)&62.5$\pm$0.9 (16.17)&1.21\\
6$^3$&279.221652 (53.20$^s$) & 38.793857 (47$^\prime$37.89\arcsec)&N/A$^5$&329.5$\pm$1.0 (14.37)&N/A\\
7$^3$&279.223982 (53.76$^s$) & 38.798338 (47$^\prime$54.02\arcsec)&273.5$\pm $0.43 (14.57)&78.9$\pm$0.9 (15.92)&0.08\\
8&279.231673 (55.60$^s$) & 38.799461 (47$^\prime$58.06\arcsec)
&86.3$\pm $0.24 (15.82)&20.1$\pm$0.8 (17.40)&-0.15\\
\hline
\enddata
\tablecomments{$^1$Positions (Epoch 2023.633) with respect to $\alpha=18^h\,36^m\,56.734^s$ and $\delta=38^o\,47^\prime\,8.121$\arcsec. $^2$Estimated color uncertainty and a possible systematic blue bias of $\sim$ 0.1 mag is discussed in the text. $^3$Source associated with Gaia DR3 object. $^4$Extended source not included in position averages. $^5$Not detected due to detector edge.}
\end{deluxetable*}

\subsection{Point Sources Interior or Close to the Debris Disk}\label{sec:pointsources}

{\color{black} Figure~\ref{fig:SUBimages} shows the PSF-subtracted images for the close-in region of Vega.} These contain the subarray science frames and were reduced using the reference frames from both sub-array and full-array; the full-array science frames do not add to the sensitivity in the speckle dominated region.

We examined the entire region interior and just exterior to the debris ring, which has a radius of $\sim$20\arcsec\ = 150 au, searching for objects using a Gaussian-smoothed image to search for $>3\sigma$ candidates. These candidates were examined visually to identify and reject stellar diffraction and other image artifacts. A primary indicator that a source may be real is the presence of the two negative lobes that result from ADI. Indeed, the ADI roll-subtraction with two rolls imprints two negative copies of the source PSF separated by the roll angle in the final image. The separation and brightness of these lobes in the data are confirmed to be real with an MCMC analysis using a forward model as discussed in the Appendix.

{\color{black}Outside the speckle-dominated region (Figure~\ref{fig:schematic})} we identified two sources (\textit{S1} and \textit{S2}) approximately 6\arcsec\ (46 au) away from Vega (Table~\ref{tab:sources}, Figure~\ref{fig:NIRCamMCMC}). Both appear to be extended. {\color{black} The first source is also identified with an object seen in MIRI data ($\S$\ref{sec:close}). A third, marginal 2.5-3$\sigma$ detection is described in the Appendix (Figure~\ref{fig:S3}, Table A1).}

\subsection{Sources Within 60\arcsec\ of Vega in the Full Frame Image}
We examined the full frame images starting with the F444W data which is more sensitive than the F210M data due to the wider bandwidth of the filter. The image was smoothed with a $\sigma$=3 pixel (FWHM=7 pixels, 0.48\arcsec) and searched for objects above 5-$\sigma$ which is approximately the limit for sources visible by naked eye inspection. This initial sample was examined visually for artifacts, obviously extended objects or locations close to the edge of the detector frames. Aperture photometry was performed on sources within 60\arcsec\ of Vega ($<$ 465 au). \textit{Astropy} routines were used on the unsmoothed image with a beam radius of 4 pixels (0.25\arcsec) and a background annulus of 16-24 pixels (1.6\arcsec-1.9\arcsec). 
The coordinates of the F444W sources were used as seeds to identify sources in the unsmoothed F210M image. After re-centering the F210M apertures, aperture photometry was carried with the same 0.25\arcsec\ aperture.

The F444W coordinates obtained using the pipeline WCS keywords showed an offset from the expected position of Vega and the Gaia objects visible in the image (Figure~\ref{fig:NIRCamSources}). An offset of $(\Delta\alpha,\Delta\delta)=(234\pm46,292\pm35)$ mas was determined with respect to the location and symmetry of the occulted image of Vega and of 8 Gaia objects. An offset of this magnitude arises from uncorrected distortions through the coronagraphic wedge and at the coronagraphic mask's location near the edge of the NIRCam field (M. Perrin and J. Girard, private communication). Future versions of the NIRCam pipeline will be updated with the latest distortion model to mitigate this effect for the coronagraphic modes. The derived offset has been applied to the positions listed in Table~\ref{tab:sources2}. The positional dispersion of an individual source is $\sim$ 100 mas resulting from uncorrected distortion in the roll-subtracted images and from the extended nature of a number of the objects seen through the complex coronagraphic PSF. 


The 0.25\arcsec\ aperture size is large relative to JWST's nominal resolution, but the diffraction from the coronagraphic mask and Lyot stop results in a much broader PSF, with a large fraction of the flux dispersed to wider angles; only 20\% and 25\% of the flux is enclosed within the aperture for the F210M and F444W filters, respectively. We simulate the coronagraphic PSF with \texttt{WebbPSF\_ext} \citep{Leisenring2024}, deriving aperture corrections of 2.87 and 4.5 
(multiplicative factors on the flux) for F210M and F444W, respectively, for our chosen aperture. We determined the photometric uncertainty by comparing the standard deviation of the fluxes in the background annulus with the integrated signal within the source aperture. 
The measured fluxes and uncertainties are given in Table~\ref{tab:sources2}.
It should be noted a number of these objects are slightly extended so that the photometry in the 0.25\arcsec\ aperture and through the complex PSF of the coronagraphic mask is unlikely be more accurate than 0.1 mag. The slightly blue color for some of the objects in Table~\ref{tab:sources2}, ([F210M]-[F444W]$\sim$-0.1 mag), suggests that the aperture correction derived from \texttt{WebbPSF\_ext} is a slight overestimate. An empirical derivation of the aperture correction for 4 stellar Gaia sources is consistent with a slightly smaller aperture correction.

\subsection{Limits to Disk Emission\label{sec:disk}}

Although Vega has a prominent debris disk seen in thermal emission, detection of a visible light counterpart has proven elusive particularly as the disk is nearly face-on. A deep HST program to search for scattered light around Vega (PID 16666) shows tentative evidence for scattered light signal from $\sim$ 10.\arcsec 5 - 30\arcsec\ with a peak surface brightness $< 2 $ MJy sr$^{-1}$ and a relatively shallow radial profile \citep{Wolff2024}. Detection of a disk in the near-IR is more challenging than at visible wavelengths due to the lower stellar flux ($f_\nu\propto \lambda^{-2}$) and typically smaller scattering cross-sections for small grains ($\lambda^{-4}$). To search for scattered light we used a PSF-subtracted image (Figure \ref{fig:VegaDisk}a), computed via a simple RDI, similar to what is described in Sec.~\ref{PSFsubtraction} but without using the other roll to avoid self-subtraction of the disk. The 5 available reference frames (from the 5-POINT dither pattern described in Sec.~\ref{sec:observations}) were combined in a linear combination, the coefficients of which were computed by maximizing the speckle subtraction in the inner 100$\times$100 pixel around the center of the star. 

Figure \ref{fig:VegaDisk}a shows the resultant F444W image along with arc-shaped apertures lying between the primary diffraction spikes used to measure the surface brightness as a function of separation from Vega. To eliminate the effects of point sources, emission with an absolute value greater than 1 MJy sr$^{-1}$ was masked out (positive values from the Vega image or negative ones from the reference star image). Figure \ref{fig:VegaDisk}b shows the median surface brightness values in the PSF-subtracted image in each annular arc after subtraction of a background determined within the dotted square shown in the image. Apart from a slight negative offset and an artifact around 10\arcsec\ visible in the image and in the background trace, there is no evidence for any disk emission at the level of $\sim$0.5 MJy sr$^{-1}$ (3$\sigma$).

\section{Results\label{sec:results}}

\subsection{Sources Close to Vega\label{sec:close}}

\begin{figure*}[t!]
\centering
\includegraphics[width=0.49\textwidth]{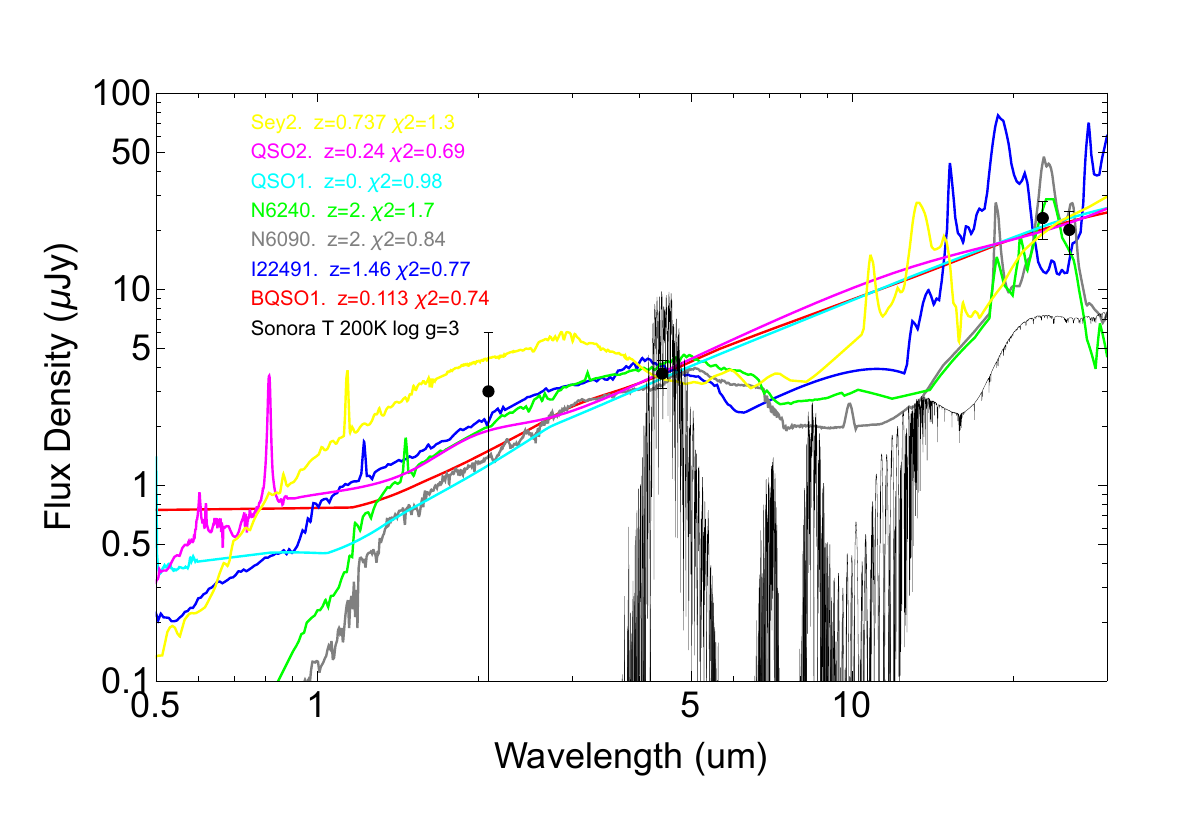}
\caption{The three JWST data points (1 NIRCam, 2 MIRI for the source \textit{S1} are compared with illustrative SEDs of a number of extragalactic objects (from a suite of Spitzer SWIRE templates \citet{Poletta2007}) and with the spectrum of cloudless T$_{eff}$=200 K, log(g)=3 Sonora Bobcat model \citep{Marley2021}. \label{fig:Sonora}}
\end{figure*}

Table~\ref{tab:sources} gives properties of the two F444W sources located close to or interior to the disk revealed in the MIRI data. Sources $S1$ and $S2$ appear extended and thus extragalactic in nature. 

Object $S1$ is detected as a point source in F444W and both MIRI F1550W and F2550W filters (Figure~\ref{fig:NIRCamMIRI}). However, its Spectral Energy Distribution (SED) has a much larger ratio of MIRI brightness to F444W than expected from representative exoplanet models (Figure~\ref{fig:Sonora}). The SED suggests it is a background galaxy similar to those identified in the vicinity of Fomalhaut \citep[Figure~\ref{fig:Sonora}]{Gaspar2023, Ygouf2023, Kennedy2023}. As discussed in \citet{Ygouf2023} we fit the limited data for this object (1 NIRCam point and two MIRI points) to a suite of Spitzer SWIRE templates \citep{Poletta2007}. The template SEDs cover visible to far-IR wavelengths for 25 galaxy types, including ellipticals, spirals, AGN, and starburst systems\footnote{http://www.iasf-milano.inaf.it/~polletta/templates/swire\_templates.html}. The existing data are not highly constraining, ruling out elliptical and late type spirals with little or no star formation. A wide variety of IR-bright galaxies and AGN at redshifts from 0.5-1.5 are consistent with the existing data. Also shown is a Sonora model \citep{Marley2021} for a cold, low luminosity planet (T$_{eff}$=200 K and log g=3) consistent with the 700 Myr age of the system. The MIRI emission of the detected object is almost a factor of 10 above that of the model.

There are no other NIRCam sources at the level of $\sim$1 \mj\ found within the disk extent ($\sim$15--25\arcsec). If the observed ALMA structures are due to a sculpting planet, then the predicted minimum mass of $\sim$6 \me\ (0.02 \mj) \citep{Matra2020} is well below the our current mass limit (0.3-0.5\mj\ or 100-150 \me).

While sources $S1$ and $S2$ are likely extragalactic in nature it is instructive to consider the appearance of a planet of comparable F444W brightness. From the age of Vega an object with an F444W brightness of [F444W]=19.7 mag would have a mass corresponding to a $<$0.5 \mj\ planet using Ames-Cond, BEX, Sonora, Spiegel \& Burrows models \citep{Baraffe2003,Spiegel2012,Linder2019,Marley2021}. Figure~\ref{fig:Sonora} shows Sonora models with (T=200 and 250 K, log g=3) corresponding roughly to a $\lesssim$0.5\mj\ planet with an age of 700 Myr, normalized to the observed F444W flux density. The figure also shows the SED of \textit{S1} --- presumably a distant galaxy (as discussed above).

As noted in \citet{Ygouf2023} contamination by extragalactic sources is always an issue with imaging at JWST's sensitivity. The incidence of extragalactic sources with brightness of F444W $\sim$19 Vega mag, is expected to be 15 sources per sq.~arcmin \citep{Ashby2013,Ygouf2023}. The area of the Vega disk is approximately $\pi (15\arcsec)^2=700^\square$\arcsec $\sim 0.2^\square\arcmin$ so that we might expect $\sim$3 background galaxies within the disk. Thus the discovery of at least two extragalactic objects within the extent of the Vega disk is not surprising.

\subsection{Other Sources}

\begin{figure}[t!]
\centering
\includegraphics[width=0.5\textwidth]{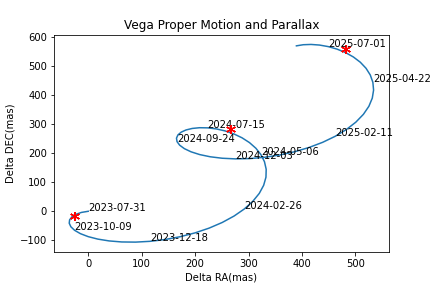}
\caption{The figure shows the combined effects of proper motion and parallax of Vega as a function of time since NIRCam's initial observation on 2023-Aug-19. The initial epoch and two future epochs within JWST's availability window are marked. Even a year after these initial observations the position offset relative to the initial observation will be $>250$ mas and easily detectable. \label{fig:VegaPM}}
\end{figure}

Of the many sources visible in the full frame images shown in Figure~\ref{fig:NIRCamSources}, eight located within 60\arcsec\ (450 au) are listed in Table~\ref{tab:sources2}. Four have counterparts in Gaia and are not physically associated with Vega because their proper motion has carried it over 2\arcsec\ in the 7 years between the Gaia and current epochs (Figure \ref{fig:VegaPM}). All but one have counterparts at F210M with colors [F210M]-[F444W]$<$1.5 mag and are thus consistent with typical stellar or extra-galactic objects \citep{Ygouf2023}. None of them have colors suggestive of low mass planet candidates which at this age (700 Myr) and brightness $14<[F444W]<18$ mag would have a very red color (F210M]-[F444]$>$6 mag \citep{Linder2019}. The one red object in Table (\#5 with [F210M]-[F444]=1.2) fits Sonora model objects with effective temperatures $\sim 1500$K, but such an object is predicted to be much brighter ($8<$[F444W]$<10$ mag) than the observed brightness of \#5 (F444W=16.2 mag). An extragalactic object \citep{Poletta2007} or a distant L/T dwarf is a more likely counterpart \citep{Hainline2020}.

\subsection{Limits to Scattered Disk Light }

The detection of scattered light around Vega is challenging due to a number of factors. First, the disk will be close to face-on which minimizes the integrated optical depth. Second, at NIRCam wavelengths, the stellar brightness is fainter and the scattering cross sections typically lower than in the visible. To compare the NIRCam results with the visible detection (or limit; \citet{Wolff2024}), we approximate the scattered light surface brightness, $I(R)$ at a distance $R$ from Vega as:

\begin{equation}
 I(R)\simeq \frac{F_{\nu,*}}{ \Omega_R}\tau_{sca}
\end{equation}

\begin{figure*}[t!]
\centering
\includegraphics[width=0.48\textwidth]{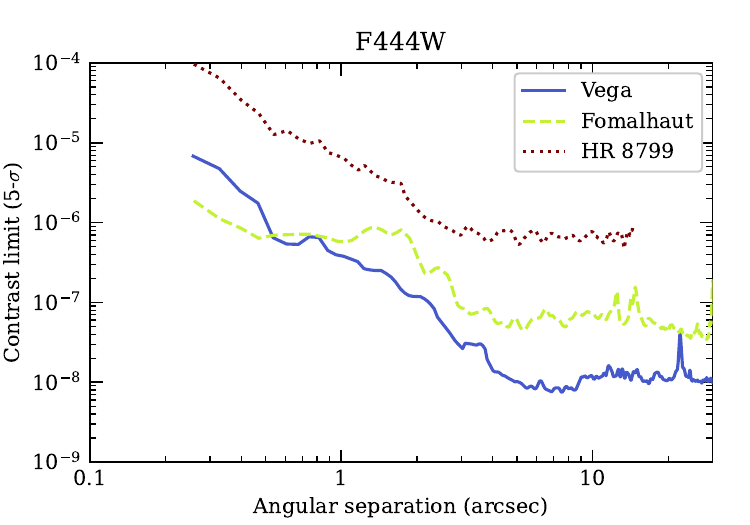}
\includegraphics[width=0.48\textwidth]{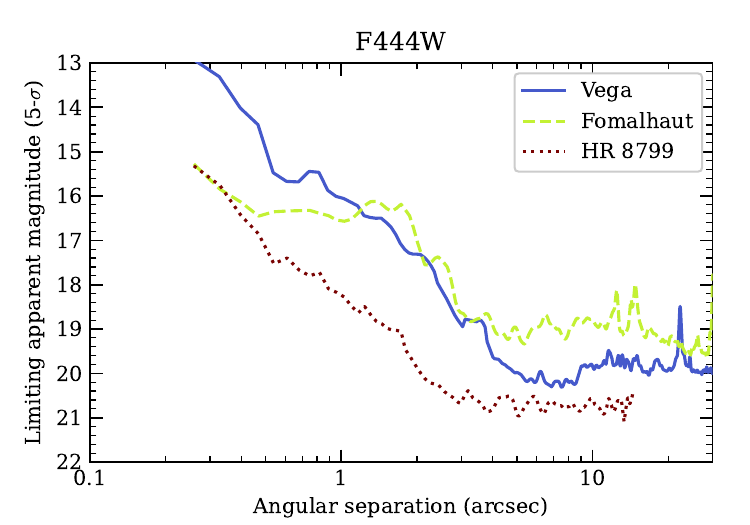}
\includegraphics[width=0.48\textwidth]{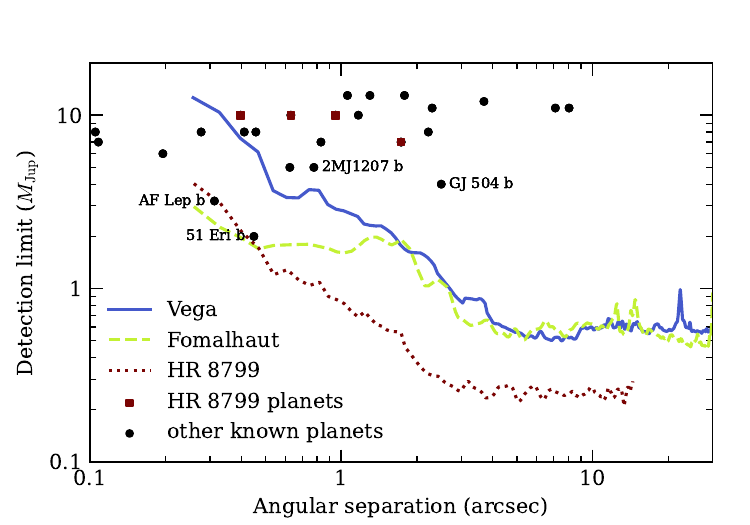}
\includegraphics[width=0.48\textwidth]{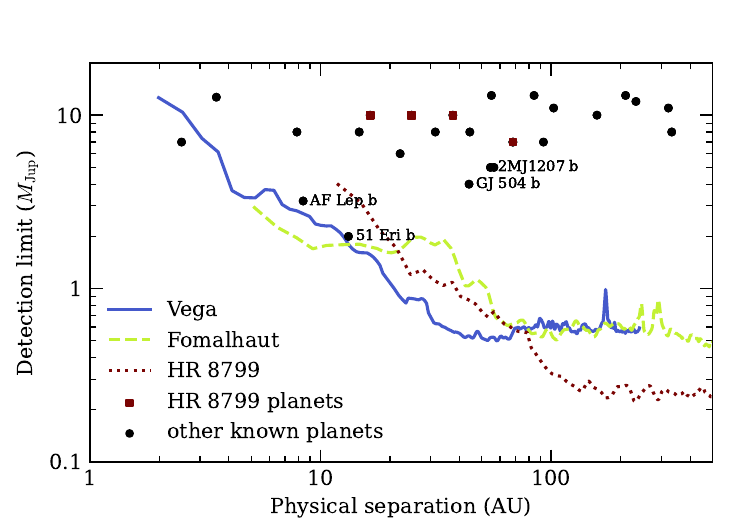}
\caption{ top, left) Contrast curves ($5\sigma$) for Vega, Fomalhaut \citep{Ygouf2023} and HR 8799 \citep{Bryden2024}. The differences are consistent with the brightness of the host star and the observation parameters as discussed in the text. top, right) Limiting magnitudes toward each star. bottom, left) Limiting mass sensitivity curves as a function of angular separation for the three stars using BEX models \citep{Linder2019}. A selection of directly imaged planets with model masses $<$13\mj\ from the NASA Exoplanet Archive are plotted as black circles; the four HR 8799 are highlighted as red squares. For HR 8799 the region interior to HR 8799 b (1.7\arcsec\ $\sim$80 AU) is omitted to recognize the four high mass planets found interior to this separation. The bump in the Vega curve at 175 AU shows the edge of the coronagraphic mask. bottom, right) Same as for (bottom, left) but plotted as a function of physical separation. The mass curves and the points for HR 8799 are corrected for projection effects -- $5\arcdeg$ inclination for Vega \citep{Monnier2012}, $67\arcdeg$ for Fomalhaut \citep{Gaspar2023}, and $26\arcdeg$ for HR 8799 \citep{Matthews2014}.
\label{fig:MassSensitivity}}
\end{figure*}

\noindent where $\Omega_R=\pi\frac{R^2}{d^2}$, $\tau_{sca}$ is the scattering optical depth, $d$ is the distance to the star from Earth, and $F_{\nu,*}$ is the measured flux density of the star at F444W. For the $I<$0.5 MJy sr$^{-1}$ limits estimated here (Figure~\ref{fig:VegaDisk}), the limits to the optical depth at 10\arcsec-20\arcsec\ are $\tau<2-8\times 10^{-5}$ which can be compared to the tentative HST detection (Wolff et al 2024, submitted) which suggests $\tau_{sca}\sim 7\times 10^{-6}$ at visible wavelengths. The MIRI-determined emission optical depths are $\tau_{em}=1-2 \times 10^{-5}$ at comparable separations \citep{Su2024}. For small ``astronomical" silicate grains, $a=$ 0.4 (0.16) $\mu$m, the ratio of absorption to scattering cross-sections is $\frac{Q_{abs}}{Q_{sca}}\sim 4 \, (68)$ \citep{Draine1984}, so that the emission optical depth determined from the MIRI data would imply a scattering optical depth of $\tau_{sca}=2.3 \, (0.15)\times 10^{-6}$ at F444W. Smaller grains would lead to considerably smaller predicted scattering optical depths, relative to the upper limits measured here of a few $\tau_{sca}=10^{-5}$. Similar arguments apply even if the grains are composed of ices rather than by astronomical silicates as has been suggested to explain HST scattered light observations \citep{Wolff2024, Tazaki2021}. Thus, on the basis of the HST and MIRI data and typical grain properties, the failure to detect scattered light at F444W is not surprising.
\section{Discussion\label{sec:discuss}}
\subsection{Planets and the Structure of the Debris Disk}

The role of planets in sculpting debris disks is a matter of active theoretical and observational investigation with a wide range of planet masses and locations identified as plausible shepherds for disk structures \citep{Chiang2009, Pearce2024}. The dust emission revealed in the MIRI images \citep{Su2024} is extremely smooth with only a modest dip in brightness in the $\sim$5\arcsec-10\arcsec (40-70 au) region. While the NIRCam source ($S1$) is located within this region its colors suggest to be an unrelated extragalactic object. There is no NIRCam object corresponding to a mass $\geq$ 0.5\mj\ associated with the MIRI gap (Figure ~\ref{fig:detectionLimits}). Nor are there objects of higher masses ($>$4\mj) detected interior to the gap or at comparable masses at any exterior point region defined by the F2550W or ALMA disks ($<$20\arcsec) which cannot be attributed to likely background objects. 

This dearth of relatively massive planets out to 200 au is consistent with the MIRI disk images \citep{Su2024} which suggest that planets within the debris system and with masses well below one \mj\ would cause detectable, but not seen, perturbations in the dust distribution. The smoothness of the disk imaged by MIRI argues against the existence of any planets with masses $>$ \mj\ as such an object would likely create structure observable in the disk image.

\subsection{Is There a Low Mass Cutoff to Planet formation?}

The presence of planets on distant orbits (10s to 100s of AU) presents a number of theoretical challenges. Planet formation via core accretion relies on the gravitational infall of gas in the protoplanetary nebula onto a solid (rocky) core of some 10s of \me. Demographic data for FGK stars suggest that this process peaks inward of 5 AU \citep{Fulton2021}. Less is known about the demographics of planets orbiting A stars as Radial Velocity (RV) data are unavailable, but it is likely that a larger, more massive disk might produce planets at larger orbital separations \citep{Mordasini2012}. The effects of migration or planet-planet scattering can put core accretion planets onto wide (but likely highly eccentric) orbits \citep{Vorobyov2013,Izidoro2023}.

The alternative formation scenario, direct collapse of nebular material into objects with masses of a few \mj\ has two variants: a) formation via gravitational collapse of two clumps of gas within a single natal molecular cloud, \textit{cloud collapse}, to create a bound, large mass ratio, binary system \citep{deFurio2022,Calissendorff2023}; or b) formation via gravitational collapse of material within the protostellar disk of the host star, \textit{disk fragmentation}, which can explain the formation of gas giants on 30-50 au orbits \citep{Boss2024,Gratton2024}. The dearth of low mass brown dwarf companions suggests that \textit{cloud collapse} is less likely than \textit{disk fragmentation} as the source of most gas giant planets on wide orbits. Recently results from JWST and EUCLID suggest that both mechanisms are robust in the field and in dense cluster environments.Using JWST \citet{Pearson2023} found over 500 Planetary Mass Objects (PMOs), many with masses 0.7- 1 \mj, toward the Trapezium Cluster in Orion; a significant fraction of these are in near-equal mass binary systems suggesting \textit{disk fragmentation}. Using early release EUCLID data \citet{Martin2024} found no low-mass cutoff of the initial mass function for free floating objects in the $\sigma$ Ori cluster down to $\sim$4 \mj. 

It is within this context that we explore the power of JWST observations like those presented here and elsewhere, e.g. \citet{Carter2023}, to probe the limits of these theories by taking advantage of JWST's sensitivity to sub-Jovian gas giants $<$ 1 \mj. 

\subsubsection{Summary of Contrast and Mass Limits}

The four panels of Figure~\ref{fig:MassSensitivity} compare contrast, apparent magnitude limits and model planet mass limits for three of the stars in the program \#1193, Vega, Fomalhaut \citep{Ygouf2023} and HR 8799 \citep{Bryden2024}. The top left panel compares the contrast limits ($5\sigma$) achieved for each star. The differences between the curves are explained by the effects of the brightness of the host star (K$\sim$0.1, 1.0 and 5.2 mag, respectively), the coronagraphic mask (MASK430R for Vega and Fomalhaut and MASK335R for HR 8799), integration times (2$\times$2509s for Vega, 2$\times$1660s for Fomalhaut, and 2$\times$936s for HR 8799 for the two rolls) and the specific detector parameters used, e.g. the number of groups. The top right panel estimates the limiting apparent magnitudes achieved toward each star. Outside of 10\arcsec, the sensitivity flattens out reaching impressively low values limited by JWST's low thermal background. The bottom two panels convert the achieved contrast levels into mass limits using COND models as a function of physical and angular separation (left and right, respectively).

\begin{figure*}[t!]
\centering
\includegraphics[width=0.7\textwidth]{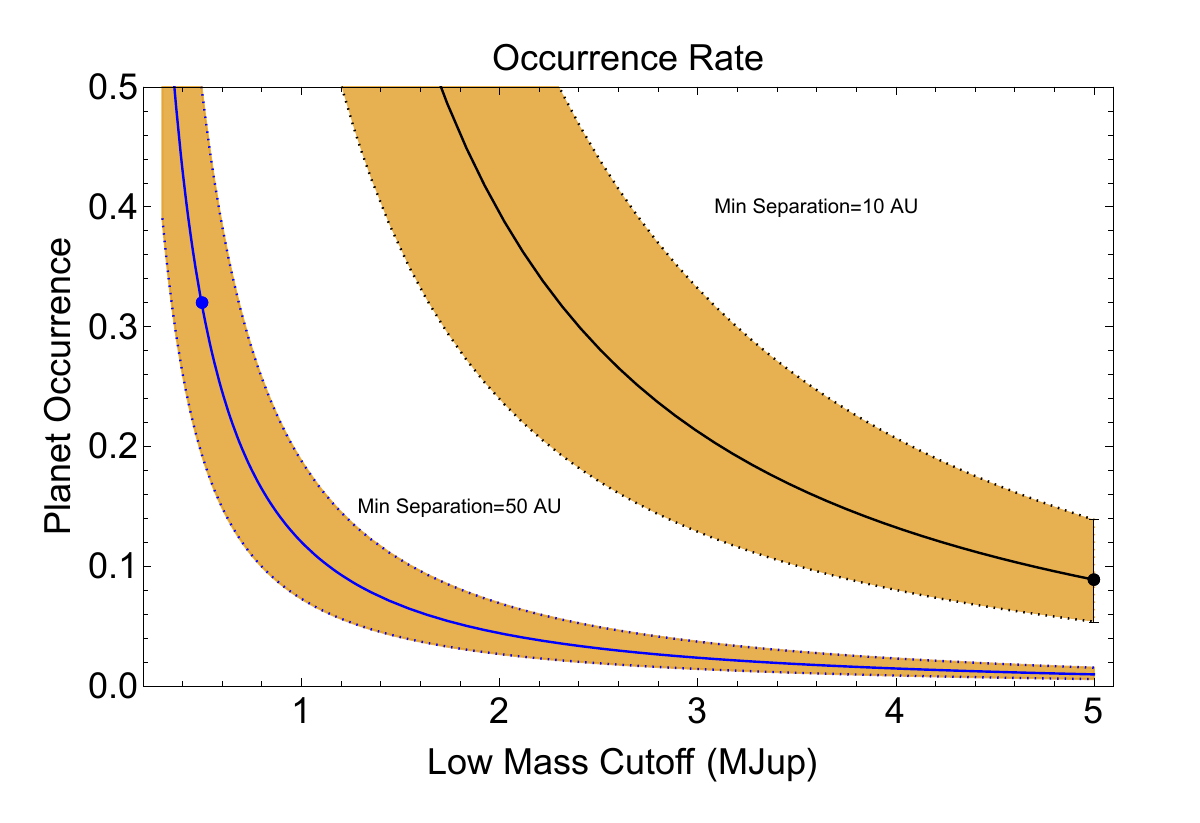}
\caption{ The planet occurrence rate for a planet in two separation ranges (10-100 AU (black) and 50-100 AU (blue)), as a function of the lower mass cutoff, $M_{min}$, integrated up to to 13 \mj, following the formulation of Eqn~\ref{prob}. The \citet{Nielsen2019} occurrence rate of 8.9\% (10-100 AU, 5-13 \mj) is indicated as a black data point while the predicted occurrence rate of $\sim$30\% at the NIRCam sensitivity level of 0.5-13 \mj\ for 50-100 AU is indicated by the blue data point. The yellow shaded regions surrounding each line denote the $\pm 1\sigma$ uncertainty due to the uncertainty in the the value of the nominal occurrence rate, $f$\label{fig:Cutoff}}
\end{figure*}

The NIRCam observations achieve a limiting mass sensitivity of 0.5 \mj\ for separations outside of $\sim$ 50 au (Figure~\ref{fig:MassSensitivity}a) and extending well beyond 100 au. Between HR 8799, Fomalhaut and now Vega we find no new planets with NIRCam despite an increase of a factor of almost 10 in mass sensitivity compared to earlier searches \citep{Nielsen2019,Vigan2021}. While acknowledging that the current sample is small, it is worth considering how JWST will contribute to our understanding to the formation of sub-Jovian mass objects. 


\citet{Nielsen2019} fit the occurrence of massive planets around stars of Vega mass. They derive a power law formulation for the frequency of planets in the range 10-100 au and 5-13 \mj:

\begin{equation}
\frac{dN^2}{dm\ da}=f C m^\alpha a^\beta
\label{prob}\end{equation}

\noindent with $f=8.9^{+5}_{-3.6}$\%, a planet mass index $\alpha=-2.37\pm 0.8$ and an orbital separation, $a$, index of $\beta=-1.99\pm0.5$. $C$ is a constant \citep[(their Eqn. 3)]{Nielsen2019} normalizing over the mass and semi-major axis range. The mass and separation indices are both quite steep implying that a change in either one would have a profound effect on the predicted occurrence rate. The uncertainties in the indices are large with other estimates as low as $\alpha=-1.3$ \citep{Cumming2008}. However, in the region of overlap with RV studies the \citet{Nielsen2019} and RV distributions (5-13 \mj\ and 10-100 au) are similar \citep[their Figure 9]{Fulton2021}.

 If we extend this distribution down to 0.5 \mj\ for the full range of separations (10-100 AU; black curve in {\color{black} Figure~\ref{fig:Cutoff}), } then the expected frequency increases to unity probability of finding a planet at this mass sensitivity level. However, NIRCam achieves its best mass sensitivity only outside of 10\arcsec\ ($\sim$50 au depending on the distance to the target; Figure~\ref{fig:MassSensitivity}) and the expected increase in the planet occurrence rate is about a factor of 4, from 8.9\% to $\sim$30\% (lower, blue curve in Figure~\ref{fig:Cutoff}). 

 We recognize that the failure to find planets around just two stars, Vega and Fomalhaut (or three including the lack of new planets exterior to the known four in HR 8799, \citet{Bryden2024}) has little statistical significance. But as additional observations of similar systems accumulate using NIRCam and MIRI, than the sample will become large enough to assess whether there is a low mass cutoff around 1 \mj\ for objects formed at large distances via gravitational instability.

There are theoretical and observational grounds to expect this cutoff. Theory suggests a lower mass limit, the Jeans Mass, where the thermal pressure and other dynamical effects overpower gravitational attraction and a clump of gas dissipates before it can collapse \citep{Low1976, Boss1997}. Recent 3-D hydrodynamic models suggest the lower limit at which self-gravitating clumps might turn into planets have Jeans masses $ \geq 0.5$ \mj\ \citep{Boss2021}. The models of \citep[their Figure 4]{Vigan2021} show a similar dearth of low mass planets formed via gravitational instability on wide orbits.

Similarly, Figure 4 of \citet{Vigan2021} shows a low occurrence rate for core accretion objects exterior to 10 au, making the region explored by NIRCam somewhat of a wasteland. It is worth pointing out, however, that one of the primary arguments in favor of the existence of planets at these large separations is the complex disk structure revealed by ALMA \citep{Andrews2018} (and to a lesser extent by MIRI) with gaps, dynamical kinks or other structures. These gap-inducing planets have a broad range of predicted masses mass \citep[$0.01<M<10$\mj]{Lodato2019}, many of which would be undetectable in the present study (M$\sim$0.5\mj). Putting planets, formed by either core accretion or gravitational instability and moved hither and yon via migration, at these large separations presents a theoretical challenge. Surveys of more systems with JWST's great sensitivity will address the ability to form sub-Jovian planets at these large separations,especially beyond 100 AU where population synthesis models suggest an increased occurrence rate due to gravitational instability \citep[their Fig. 4]{Vigan2021}.

\section{Conclusions\label{sec:conclude}}

We have observed the original debris disk star, $\alpha$ Lyra, using the NIRCam coronagraph. Three sources are detected at F444W within the area defined by the bright debris disk imaged by MIRI. It is likely that two objects are extragalactic in nature. A third marginally detected source has no counterparts at F210M or in the MIRI data. Its nature is uncertain and astrometric confirmation will be required to associate it with Vega (Figure~\ref{fig:VegaPM}). If associated with Vega it would have a mass of $1\sim$3 \mj\ and an effective temperature of 250K. The presence of such a massive planet would be disruptive of the smooth disk structure in the MIRI data which argues against the exoplanet interpretation. 

Beyond a separation of $\sim$4\arcsec\ we achieve a contrast level of a few$\times10^{-8}$ corresponding to a mass limit of $\sim$0.5-1 \mj. A planet at or more likely somewhat smaller than this level could be responsible for the any structures in the MIRI or ALMA disks. Additional observations could push this limit to $<$0.5\mj\ into the realm of Saturn or even Uranus mass objects. 

\section{Appendix 1. Details of MCMC Analysis}\label{sec:MCMC}

\setcounter{figure}{0} 
\renewcommand\thefigure{A\arabic{figure}}

\setcounter{table}{0} 
\renewcommand\thetable{A\arabic{table}}

{The astrometry and photometry of point-like sources are estimated with an MCMC \citep[\texttt{emcee};][]{Foreman-Mackey2013} fit to the PSF-subtracted data as implemented by \texttt{pyKLIP} \citep{Wang2015}. The model includes the x and y position in the detector, a flux scaling factor to a reference flux, and a correlation length. In this process, a forward model of the NIRCam PSF is used to fit the data, which ensures that the modifications in the PSF structure due to the post-processing are accounted for. In particular, the prominent two negative lobes that result from ADI are of great importance for the MCMC process to increase confidence in the resulting posteriors. We use 100 walkers for the MCMC fit, with \textit{burn-in} 200 steps, and 500 steps after \textit{burn-in}.
The position of Vega in the detector is determined by performing cross-correlations of the data with synthetic PSFs obtained with \texttt{WebbPSF} \citep{Greenbaum2023}. We use the \texttt{chi2\_shift} functions in the \texttt{image-registration} Python package\footnote{https://image-registration.readthedocs.io/} which yields uncertainties of $\sim$7~mas, consistent with the result of \citealt{Carter2023}. The resultant fits and corner plots are shown in Figures~\ref{fig:NIRCamMCMC} and \ref{fig:S3} with derived values given in Table~\ref{tab:sources}.

\begin{figure*}[t!]
\centering
\includegraphics[width=0.75\textwidth]{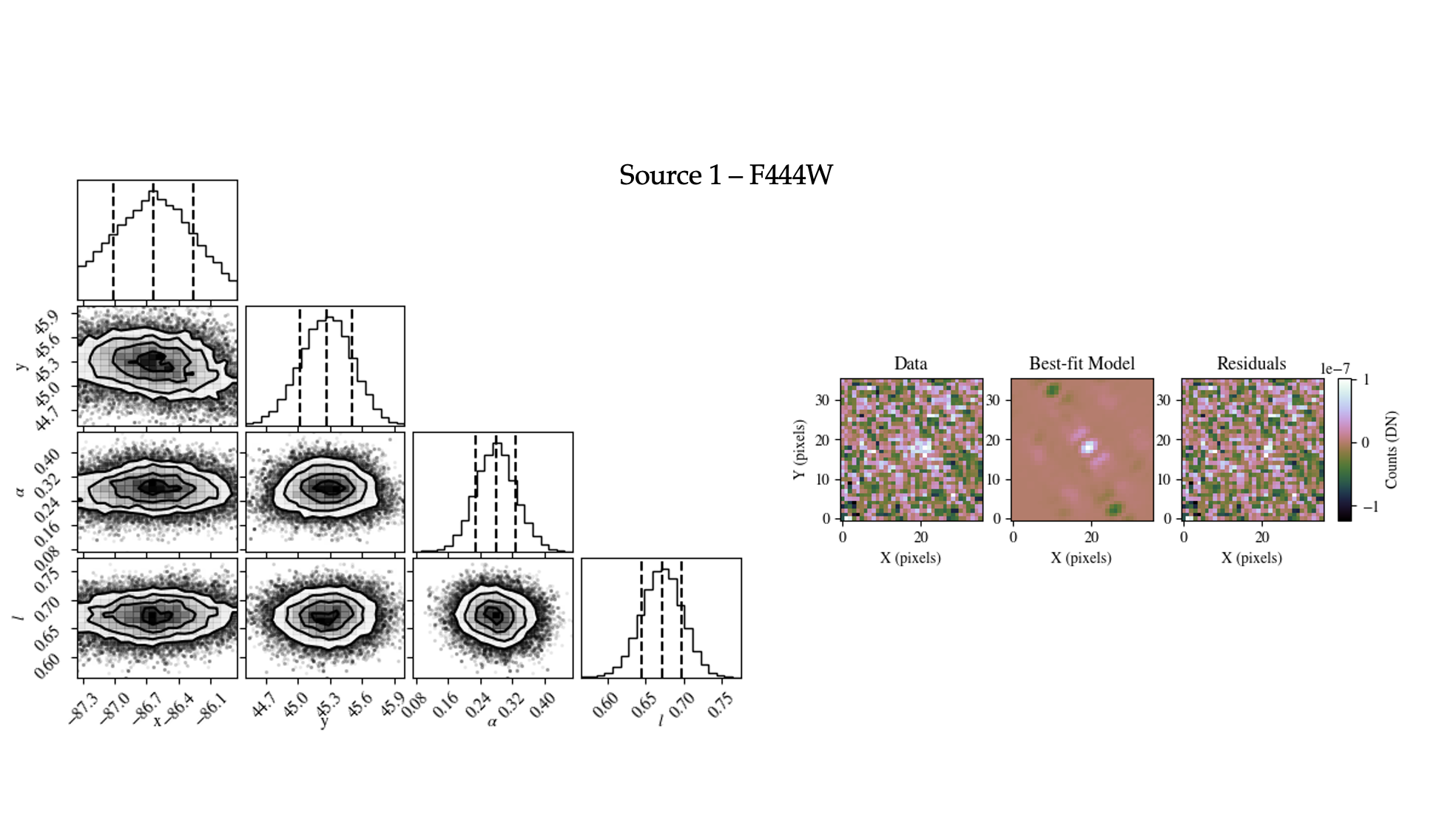} \\
\includegraphics[width=0.75\textwidth]{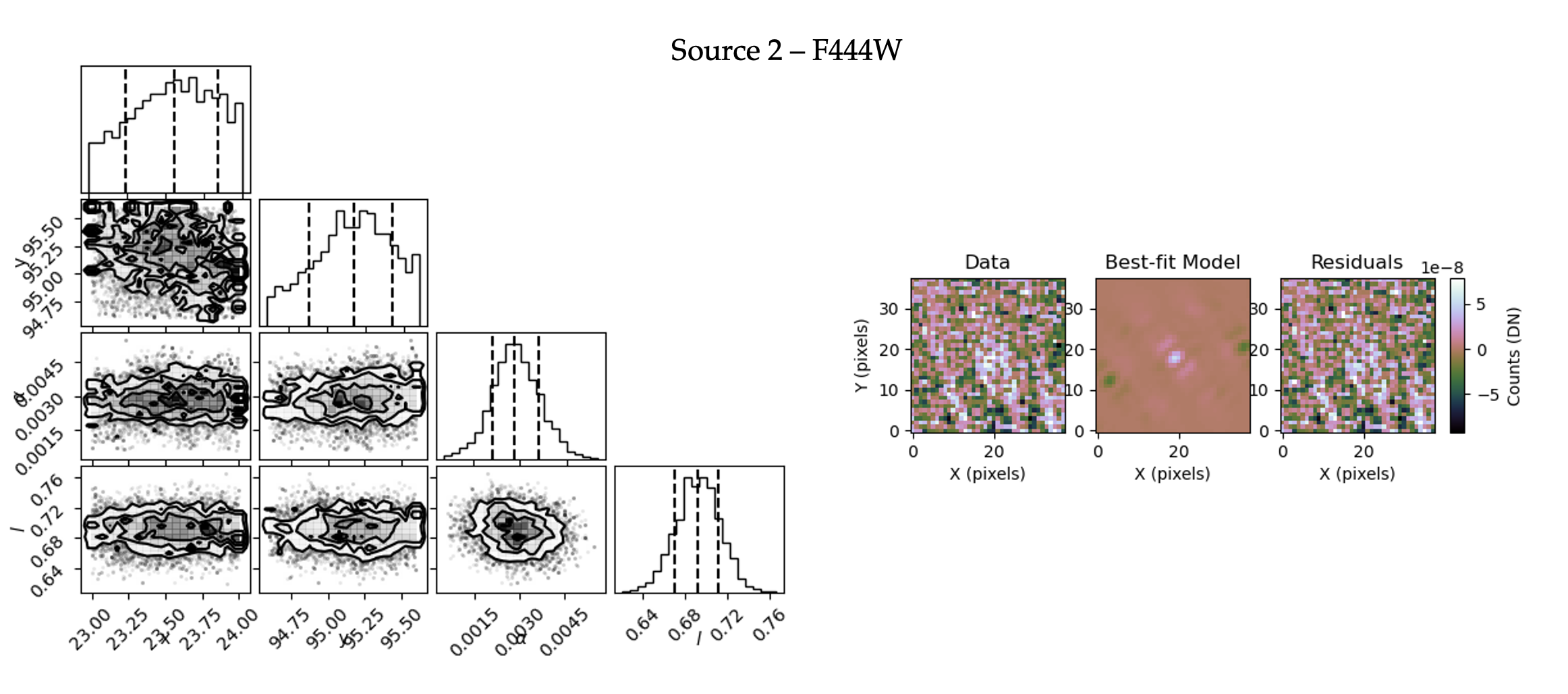}
\caption{top) MCMC analysis of the F444W source \textit{S1} located at roughly 2 o'clock position and 6\arcsec\ away from Vega (Table~\ref{tab:sources},Figure~\ref{fig:NIRCamMIRI}). bottom) MCMC analysis of the F444W \textit{S2} source located at roughly 11 o'clock position and 6\arcsec\ away from Vega. this source appears to be slightly extended.
\label{fig:NIRCamMCMC}}
\end{figure*}

\section{Appendix 2. Analyzing the Speckle-Dominated Region: A Marginal Detection close to Vega}
The region dominated by residual speckles near the star, i.e. within $\sim1.5$\arcsec, is a particularly interesting regime for point source detection. Many speckles mimic the structure that a real object would exhibit (see Sec.~\ref{sec:pointsources}).
Since there is no obvious point source that stands out with respect to the stellar speckle noise, we examine the brightest clumps of pixels to fully discard their astrophysical origin.

We compute the SNR map in the inner region, based on the aperture photometry with respect to the noise in the corresponding separation. Within 2\arcsec, 9 clumps of bright pixels show an SNR of 2.5 or more, and none are above 3.5. We discard those that are either a single pixel or don't exhibit any negative-positive-negative structure. For the four remaining blobs we perform a model fit using the MCMC process described in Sec.~\ref{sec:MCMC}. Only one bright structure yields a model that is largely consistent with a point source.


This source of marginal significance, denoted $S3$ , is located 1.4" (10.7 au) South of Vega Figure~\ref{fig:S3}. It has no counterpart at F210M or in the MIRI data. If $S3$ were associated with Vega, then its apparent magnitude of [F444W]=16.5 mag (Table~\ref{tab:S3}) would correspond to an absolute mag of 17.1 mag and a mass of approximately 2-3M$_{Jup}$ based on a Sonora Bobcat model \citep{Marley2021} with an age of 700 Myr and effective temperature of 200-250 K. Such an object would not be detectable in our F210M data.

\begin{deluxetable*}{llclllll}[t!]
\tablewidth{0pt}
\tablecaption{NIRCam Sources Found Close to Vega\label{tab:S3}}
\tablehead{&\colhead{Offset}&
\colhead{Separation} & \colhead{F444W contrast}&\colhead{F$_\nu$(F210M)}&
\colhead{F$\nu$(F444W)}&\colhead{F$\nu$(F1550W)} & \colhead{F$\nu$(F2550W)}\\
\colhead{ID}&\colhead{$\Delta \alpha,\Delta\delta$\, (\arcsec)} & \colhead{(AU)}&\colhead{($\times10^{-8})$}&
\colhead{($\mu$Jy)/(Vega mag)} &
\colhead{($\mu$Jy)/(Vega mag)} &
\colhead{($\mu$Jy)/(Vega mag)} &
\colhead{($\mu$Jy)/(Vega mag)}}
\startdata
S3&($-$0.43,$-$1.35)&10.7&$24\pm8$&N/A&45$\pm$13&$<$90 ($3\sigma$)&N/A\\
&$\pm$0.013\arcsec&&& &16.5$\pm$0.3& & \\
\enddata
\end{deluxetable*}

\begin{figure*}[t!]
\centering
\includegraphics[width=0.4\textwidth]{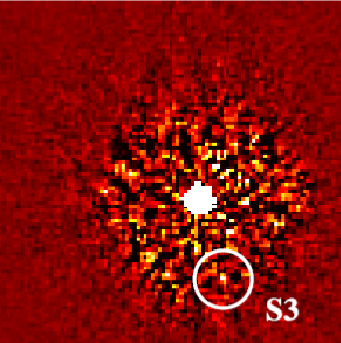} \\
 \includegraphics[width=0.45\textwidth]{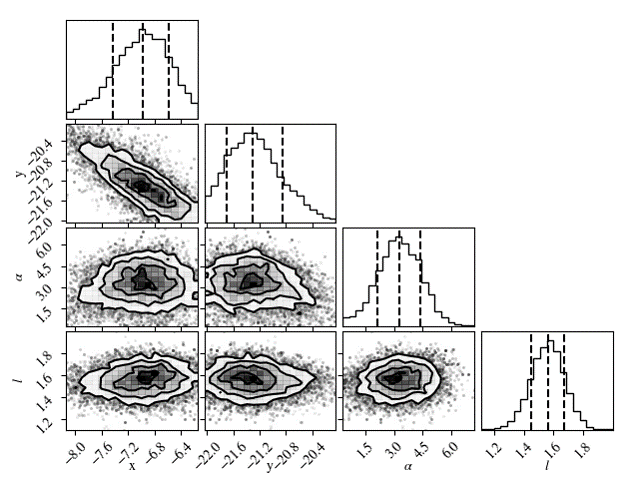}
\includegraphics[width=0.45\textwidth]{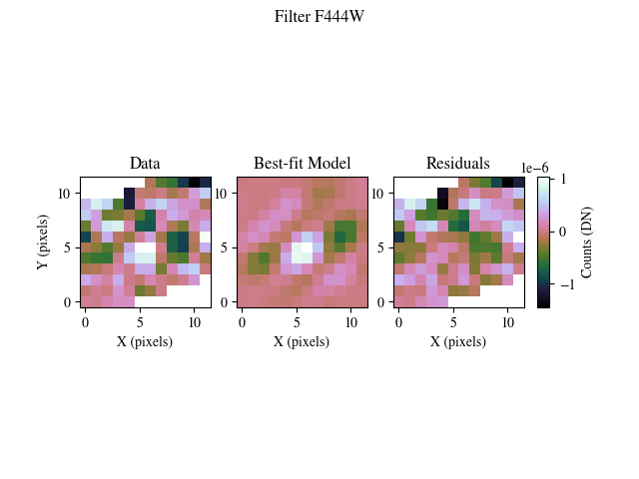}
\caption{top) An inset from Figure~\ref{fig:schematic} shows the position of the marginal detection discussed in this Appendix. bottom) MCMC analysis of the F444W source \textit{S3} approximately 1.4\arcsec\ due South of Vega (Table~\ref{tab:sources}).
\label{fig:S3}}
\end{figure*}

\newpage
\begin{acknowledgements}
 CAB dedicates this article to the pioneering IRAS team who first discovered the ``Vega Phenomenon", in particular to Fred Gillett and George Aumann, as well as to leaders of the IRAS mission, including Gerry Neugebauer, Harm Habing, Jim Houck, and Frank Low. Much of what JWST studies today, from debris disks to infrared-bright galaxies, has roots going back to that pioneering mission. CAB also wishes to thank M.\ Perrin and J.\ Girard for their rapid and thorough investigation into the astrometric offset seen in the full frame images. 
 
 NIRCam development and use at the University of Arizona is supported through NASA Contract NAS5-02105. Part of this work was carried out at the Jet Propulsion Laboratory, California Institute of Technology, under a contract with the National Aeronautics and Space Administration (80NM0018D0004). The work of A.G., G.R., S.W., and K.S.\ was partially supported by NASA grants NNX13AD82G and 1255094. D.J.\ is supported by NRC Canada and by an NSERC Discovery Grant. 
The High Performance Computing resources used in this investigation were provided by funding from the JPL Information and Technology Solutions Directorate. We are grateful for support from NASA for the JWST NIRCam project though contract number NAS5-02105 (M. Rieke, University of Arizona, PI).
\end{acknowledgements}
\copyright 2024. All rights reserved.

\facilities{JWST}

\software{
\texttt{astropy} \citep{astropy2022},
\texttt{jwst} \citep{jwst2022},
\texttt{NIRCoS} \citep{Kammerer2022},
\texttt{pyNRC} \citep{Leisenring2024},
\texttt{pyKLIP} \citep{Wang2015},
\texttt{SpaceKLIP} \citep{Kammerer2022},
\texttt{WebbPSF} \citep{Perrin2014},
\texttt{WebbPSF\_ext} \citep{Leisenring2024}
}


\clearpage

\begin{thebibliography}{99}
\expandafter\ifx\csname natexlab\endcsname\relax\def\natexlab#1{#1}\fi
\renewcommand{\bibfont}{\small}
\setlength{\itemsep}{0pt}





\bibitem[Absil et al.(2006)]{Absil2006} Absil, O., di Folco, E., M{\'e}rand, A., et al.\ 2006, \aap, 452, 237. doi:10.1051/0004-6361:20054522

\bibitem[{{Amara} \& {Quanz}(2012)}]{Amara2012}
{Amara}, A., \& {Quanz}, S.~P. 2012, \mnras, 427, 948,
 doi:{10.1111/j.1365-2966.2012.21918.x}

\bibitem[Andrews et al.(2018)]{Andrews2018} Andrews, S.~M., Huang, J., P{\'e}rez, L.~M., et al.\ 2018, \apjl, 869, L41. doi:10.3847/2041-8213/aaf741


\bibitem[Ashby et al.(2013)]{Ashby2013} Ashby, M.~L.~N., Willner, S.~P., Fazio, G.~G., et al.\ 2013, \apj, 769, 80. doi:10.1088/0004-637X/769/1/80

\bibitem[{{Astropy Collaboration} {et~al.}(2022){Astropy Collaboration},
 {Price-Whelan}, {Lim}, {Earl}, {Starkman}, {Bradley}, {Shupe}, {Patil},
 {Corrales}, {Brasseur}, {N{\"o}the}, {Donath}, {Tollerud}, {Morris},
 {Ginsburg}, {Vaher}, {Weaver}, {Tocknell}, {Jamieson}, {van Kerkwijk},
 {Robitaille}, {Merry}, {Bachetti}, {G{\"u}nther}, \& {Astropy Project Contributors}}]{astropy2022}
{Astropy Collaboration}, {Price-Whelan}, A.~M., {Lim}, P.~L., {et~al.} 2022,
 \apj, 935, 167
 
\bibitem[Aumann et al.(1984)]{Aumann1984} Aumann, H.~H., Gillett, F.~C., Beichman, C.~A., et al.\ 1984, \apjl, 278, L23. doi:10.1086/184214

\bibitem[Baines et al.(2018)]{Baines2018} Baines, E.~K., Armstrong, J.~T., Schmitt, H.~R., et al.\ 2018, \aj, 155, 30. doi:10.3847/1538-3881/aa9d8b

{\color{black} \bibitem[Baraffe et al.(2003)]{Baraffe2003} Baraffe, I., Chabrier, G., Barman, T.~S., et al.\ 2003, \aap, 402, 701. doi:10.1051/0004-6361:20030252}



\bibitem[Boss(1997)]{Boss1997} Boss, A.~P.\ 1997, Science, 276, 1836. doi:10.1126/science.276.5320.1836

 
 \bibitem[Boss(2021)]{Boss2021} Boss, A.~P.\ 2021, \apj, 923, 93. doi:10.3847/1538-4357/ac2e05

\bibitem[Boss(2024)]{Boss2024} Boss, A.~P.\ 2024, \apj, 969, 157. doi:10.3847/1538-4357/ad4ed4

\bibitem[Bryden et al. (2024)]{Bryden2024} Bryden, G. Llop-Sayson, J., Beichman, C. et al.\ 2024 \\apj, in press.

\bibitem[Bushouse et~al.\ (2022)]{jwst2022}
Bushouse, H., Eisenhamer, J., Dencheva, N., {et~al.} 2022, JWST Calibration
 Pipeline, Zenodo, doi:10.5281/ZENODO.7038885.
\newblock \url{https://zenodo.org/record/7038885}

\bibitem[Calissendorff et al.(2023)]{Calissendorff2023} Calissendorff, P., De Furio, M., Meyer, M., et al.\ 2023, \apjl, 947, L30. doi:10.3847/2041-8213/acc86d

\bibitem[Carter et al.(2023)]{Carter2023} Carter, A.~L., Hinkley, S., Kammerer, J., et al.\ 2023, \apjl, 951, L20. doi:10.3847/2041-8213/acd93e




\bibitem[Chiang et al.(2009)]{Chiang2009} Chiang, E., Kite, E., Kalas, P., et al.\ 2009, \apj, 693, 734. doi:10.1088/0004-637X/693/1/734

\bibitem[Ciardi et al.(2001)]{Ciardi2001} Ciardi, D.~R., van Belle, G.~T., Akeson, R.~L., et al.\ 2001, \apj, 559, 1147. doi:10.1086/322345

\bibitem[Cumming et al.(2008)]{Cumming2008} Cumming, A., Butler, R.~P., Marcy, G.~W., et al.\ 2008, \pasp, 120, 531. doi:10.1086/588487

\bibitem[De Furio et al.(2022)]{deFurio2022} De Furio, M., Meyer, M.~R., Reiter, M., et al.\ 2022, \apj, 925, 112. doi:10.3847/1538-4357/ac36d4


\bibitem[Draine \& Lee(1984)]{Draine1984} Draine, B.~T. \& Lee, H.~M.\ 1984, \apj, 285, 89. doi:10.1086/162480




\bibitem[Foreman-Mackey et al.(2013)]{Foreman-Mackey2013} Foreman-Mackey, D., Hogg, D.~W., Lang, D., et al.\ 2013, \pasp, 125, 306. doi:10.1086/670067

\bibitem[Fulton et al.(2021)]{Fulton2021} Fulton, B.~J., Rosenthal, L.~J., Hirsch, L.~A., et al.\ 2021, \apjs, 255, 14. doi:10.3847/1538-4365/abfcc1



\bibitem[Gaspar \& Rieke(2020)]{Gaspar2020} Gaspar, A. \& Rieke, G.\ 2020, Proceedings of the National Academy of Science, 117, 9712. doi:10.1073/pnas.1912506117

\bibitem[G{\'a}sp{\'a}r et al.(2023)]{Gaspar2023} G{\'a}sp{\'a}r, A., Wolff, S.~G., Rieke, G.~H., et al.\ 2023, Nature Astronomy, 7, 790. doi:10.1038/s41550-023-01962-6

\bibitem[Gillett(1986)]{Gillett1986} Gillett, F.~C.\ 1986, Light on Dark Matter, 124, 61. doi:10.1007/978-94-009-4672-9\_10

\bibitem[Girard et al.(2022)]{Girard2022} Girard, J.~H., Leisenring, J., Kammerer, J., et al.\ 2022, \procspie, 12180, 121803Q. doi:10.1117/12.2629636


\bibitem[Gratton et al.(2024)]{Gratton2024} Gratton, R., Bonavita, M., Mesa, D., et al.\ 2024, \aap, 685, A119. doi:10.1051/0004-6361/202348393


\bibitem[Greenbaum et al.(2023)]{Greenbaum2023} Greenbaum, A.~Z., Llop-Sayson, J., Lew, B.~W.P., et al.\ 2023, \apj, 945, 126. doi.org/10.3847/1538-4357/acb68b

\bibitem[Heinze et al.(2008)]{Heinze2008} Heinze, A.~N., Hinz, P.~M., Kenworthy, M., et al.\ 2008, \apj, 688, 583. doi:10.1086/592100

\bibitem[JDox Documentation (2023)]{JDoxColor} {HCI PSF Reference Stars (2023)} \url{https://jwst-docs.stsci.edu/methods-and-roadmaps/jwst-high-contrast-imaging/jwst-high-contrast-imaging-proposal-planning/hci-psf-reference-stars}


\bibitem[Janson et al.(2015)]{Janson2015} Janson, M., Quanz, S.~P., Carson, J.~C., et al.\ 2015, \aap, 574, A120. doi:10.1051/0004-6361/201424944

\bibitem[Kennedy et al.(2023)]{Kennedy2023} Kennedy, G.~M., Lovell, J.~B., Kalas, P., et al.\ 2023, \mnras, 524, 2698. doi:10.1093/mnras/stad2058


\bibitem[Hainline et al.(2020)]{Hainline2020} Hainline, K.~N., Hviding, R.~E., Rieke, M., et al.\ 2020, \apj, 892, 125. doi:10.3847/1538-4357/ab7dc3
 
\bibitem[Hurt et al.(2021)]{Hurt2021} Hurt, S.~A., Quinn, S.~N., Latham, D.~W., et al.\ 2021, \aj, 161, 157. doi:10.3847/1538-3881/abdec8



{\color{black} \bibitem[Izidoro et al.(2023)]{Izidoro2023} Izidoro, A., Raymond, S.~N., Kaib, N.~A., et al.\ 2023, AAS/Division of Dynamical Astronomy Meeting, 55, 104.01}



\bibitem[Johnson et al.(1966)]{Johnson1966} Johnson, H.~L., Mitchell, R.~I., Iriarte, B., et al.\ 1966, Communications of the Lunar and Planetary Laboratory, 4, 99


\bibitem[Kalas et al.(2005)]{Kalas2005} Kalas, P., Graham, J.~R., \& Clampin, M.\ 2005, \nat, 435, 1067. doi:10.1038/nature03601

\bibitem[Kalas et al.(2008)]{Kalas2008} Kalas, P., Graham, J.~R., Chiang, E., et al.\ 2008, Science, 322, 1345. doi:10.1126/science.1166609

{\color{black} \bibitem[Kammerer et al.(2022)]{Kammerer2022} Kammerer, J., Girard, J., Carter, A.~L., et al.\ 2022, \procspie, 12180, 121803N. doi:10.1117/12.2628865}




\bibitem[Lagrange et al.(2010)]{Lagrange2010} Lagrange, A.-M., Bonnefoy, M., Chauvin, G., et al.\ 2010, Science, 329, 57. doi:10.1126/science.1187187

\bibitem[Lawler et al.(2015)]{Lawler2015} Lawler, S.~M., Greenstreet, S., \& Gladman, B.\ 2015, \apjl, 802, L20. doi:10.1088/2041-8205/802/2/L20


\bibitem[van Leeuwen(2007)]{Hipparcos2007} van Leeuwen, F.\ 2007, \aap, 474, 653. doi:10.1051/0004-6361:20078357







\bibitem[{{Leisenring}(2023)}]{Leisenring2024}
{Leisenring}, J. 2024, \apj, in prep

\bibitem[Linder et al.(2019)]{Linder2019} Linder, E.~F., Mordasini, C., Molli{\`e}re, P., et al.\ 2019, \aap, 623, A85. doi:10.1051/0004-6361/201833873


\bibitem[Lodato et al.(2019)]{Lodato2019} Lodato, G., Dipierro, G., Ragusa, E., et al.\ 2019, \mnras, 486, 453. doi:10.1093/mnras/stz913

\bibitem[Low \& Lynden-Bell(1976)]{Low1976} Low, C. \& Lynden-Bell, D.\ 1976, \mnras, 176, 367. doi:10.1093/mnras/176.2.367


\bibitem[Marley et al.(2021)]{Marley2021} Marley, M.~S., Saumon, D., Visscher, C., et al.\ 2021, \apj, 920, 85. doi:10.3847/1538-4357/ac141d

\bibitem[Marois et al.(2008)]{Marois2008} Marois, C., Macintosh, B., Barman, T., et al.\ 2008, Science, 322, 1348. doi:10.1126/science.1166585


\bibitem[Mart{\'\i}n et al.(2024)]{Martin2024} Mart{\'\i}n, E.~L., \{{\v{Z}}\}erjal, M., Bouy, H., et al.\ 2024, arXiv:2405.13497. doi:10.48550/arXiv.2405.13497


\bibitem[Matr{\`a} et al.(2020)]{Matra2020} Matr{\`a}, L., Dent, W.~R.~F., Wilner, D.~J., et al.\ 2020, \apj, 898, 146. doi:10.3847/1538-4357/aba0a4


\bibitem[Matthews et al.(2014)]{Matthews2014} Matthews, B., Kennedy, G., Sibthorpe, B., et al.\ 2014, \apj, 780, 97. doi:10.1088/0004-637X/780/1/97


\bibitem[{{Mawet} {et~al.}(2014){Mawet}, {Milli}, {Wahhaj}, {Pelat}, {Absil},
 {Delacroix}, {Boccaletti}, {Kasper}, {Kenworthy}, {Marois}, {Mennesson}, \&
 {Pueyo}}]{Mawet2014}
{Mawet}, D., {Milli}, J., {Wahhaj}, Z., {et~al.} 2014, \apj, 792, 97,
 doi:{10.1088/0004-637X/792/2/97}

\bibitem[Mawet et al.(2019)]{Mawet2019} Mawet, D., Hirsch, L., Lee, E.~J., et al.\ 2019, \aj, 157, 33. doi:10.3847/1538-3881/aaef8a

\bibitem[Meshkat et al.(2017)]{Meshkat2017} Meshkat, T., Mawet, D., Bryan, M.~L., et al.\ 2017, \aj, 154, 245. doi:10.3847/1538-3881/aa8e9a

\bibitem[Meshkat et al.(2018)]{Meshkat2018} Meshkat, T., Nilsson, R., Aguilar, J., et al.\ 2018, \aj, 156, 214. doi:10.3847/1538-3881/aae14f

\bibitem[Metchev et al.(2003)]{Metchev2003} Metchev, S.~A., Hillenbrand, L.~A., \& White, R.~J.\ 2003, \apj, 582, 1102. doi:10.1086/344750


\bibitem[Monnier et al.(2012)]{Monnier2012} Monnier, J.~D., Che, X., Zhao, M., et al.\ 2012, \apjl, 761, L3. doi:10.1088/2041-8205/761/1/L3


\bibitem[Mordasini et al.(2012)]{Mordasini2012} Mordasini, C., Alibert, Y., Benz, W., et al.\ 2012, \aap, 541, A97. doi:10.1051/0004-6361/201117350


\bibitem[Nielsen et al.(2019)]{Nielsen2019} Nielsen, E.~L., De Rosa, R.~J., Macintosh, B., et al.\ 2019, \aj, 158, 13. doi:10.3847/1538-3881/ab16e9

\bibitem[Pearce et al.(2024)]{Pearce2024} Pearce, T.~D., Krivov, A.~V., Sefilian, A.~A., et al.\ 2024, \mnras, 527, 3876. doi:10.1093/mnras/stad3462

\bibitem[Pearson \& McCaughrean(2023)]{Pearson2023} Pearson, S.~G. \& McCaughrean, M.~J.\ 2023, arXiv:2310.01231. doi:10.48550/arXiv.2310.01231



\bibitem[Perrin et al.(2014)]{Perrin2014} Perrin, M.~D., Sivaramakrishnan, A., Lajoie, C.-P., et al.\ 2014, \procspie, 9143, 91433X. doi:10.1117/12.2056689


{\color{black} \bibitem[Perrin et al.(2018)]{Perrin2018} Perrin, M.~D., Pueyo, L., Van Gorkom, K., et al.\ 2018, \procspie, 10698, 1069809. doi:10.1117/12.2313552}


\bibitem[Peterson et al.(2006)]{Peterson2006} Peterson, D.~M., Hummel, C.~A., Pauls, T.~A., et al.\ 2006, \nat, 440, 896. doi:10.1038/nature04661

\bibitem[Polletta et al.(2007)]{Poletta2007} Polletta, M., Tajer, M., Maraschi, L., et al.\ 2007, \apj, 663, 81. doi:10.1086/518113


\bibitem[Ren et al.(2023)]{Ren2023} Ren, B.~B., Wallack, N.~L., Hurt, S.~A., et al.\ 2023, \aap, 670, A162. doi:10.1051/0004-6361/202244485

\bibitem[Rieke et al.(2023)]{Rieke2023} Rieke, M.~J., Kelly, D.~M., Misselt, K., et al.\ 2023, \pasp, 135, 028001. doi:10.1088/1538-3873/acac53



\bibitem[Smith \& Terrile(1984)]{Smith1984} Smith, B.~A. \& Terrile, R.~J.\ 1984, Science, 226, 1421. doi:10.1126/science.226.4681.1421

\bibitem[{{Soummer} {et~al.}(2012){Soummer}, {Pueyo}, \&
 {Larkin}}]{Soummer2012}
{Soummer}, R., {Pueyo}, L., \& {Larkin}, J. 2012, \apjl, 755, L28,
 doi:{10.1088/2041-8205/755/2/L28}
 

\bibitem[Spiegel \& Burrows(2012)]{Spiegel2012} Spiegel, D.~S. \& Burrows, A.\ 2012, \apj, 745, 174. doi:10.1088/0004-637X/745/2/174





{\color{black} \bibitem[Su et al.(2013)]{Su2013} Su, K.~Y.~L., Rieke, G.~H., Malhotra, R., et al.\ 2013, \apj, 763, 118. doi:10.1088/0004-637X/763/2/118}


\bibitem[Su et al.(2024)]{Su2024} Su, K.~Y.~L., Gaspar, A., Rieke, G.~H., et al.\ 2024, arXiv:2410.23636. doi:10.48550/arXiv.2410.23636


\bibitem[Tazaki et al.(2021)]{Tazaki2021} Tazaki, R., Murakawa, K., Muto, T., et al.\ 2021, \apj, 921, 173. doi:10.3847/1538-4357/ac1f8c




\bibitem[Vigan et al.(2021)]{Vigan2021} Vigan, A., Fontanive, C., Meyer, M., et al.\ 2021, \aap, 651, A72. doi:10.1051/0004-6361/202038107

\bibitem[Vorobyov(2013)]{Vorobyov2013} Vorobyov, E.~I.\ 2013, \aap, 552, A129. doi:10.1051/0004-6361/201220601


\bibitem[{{Wang} {et~al.}(2015){Wang}, {Ruffio}, {De Rosa}, {Aguilar}, {Wolff},
 \& {Pueyo}}]{Wang2015}
{Wang}, J.~J., {Ruffio}, J.-B., {De Rosa}, R.~J., {et~al.} 2015, {pyKLIP: PSF
 Subtraction for Exoplanets and Disks}, Astrophysics Source Code Library,
 record ascl:1506.001.



{\color{black} \bibitem[Wolff et al.(2024)]{Wolff2024} Wolff, S.~G., G{\'a}sp{\'a}r, A., Rieke, G.~H., et al.\ 2024, \aj, 168, 236. doi:10.3847/1538-3881/ad67cb}


\bibitem[Wyatt(2008)]{Wyatt2008} Wyatt, M.~C.\ 2008, \araa, 46, 339. doi:10.1146/annurev.astro.45.051806.110525

\bibitem[Wyatt(2018)]{Wyatt2018} Wyatt, M.~C.\ 2018, Handbook of Exoplanets, 146. doi:10.1007/978-3-319-55333-7\_146

\bibitem[Ygouf et al.(2023)]{Ygouf2023} Ygouf, M., Beichman, C. A. Llop-Sayson, J., et. al., ApJ, in press. arxiv.org/abs/2310.15028

 \bibitem[Yoon et al.(2008)]{Yoon2008} Yoon, J., Peterson, D.~M., Zagarello, R.~J., et al.\ 2008, \apj, 681, 570. doi:10.1086/588550
 
\bibitem[Yoon et al.(2010)]{Yoon2010} Yoon, J., Peterson, D.~M., Kurucz, R.~L., et al.\ 2010, \apj, 708, 71. doi:10.1088/0004-637X/708/1/71



\end{thebibliography}
\end{document}